\documentclass[a4paper,12pt]{article}
\pdfoutput=1
\usepackage{hyperref}
\usepackage{epsfig}
\usepackage{amssymb}

\usepackage{setspace}

\usepackage{amsmath}
\usepackage{amssymb}
\usepackage{graphicx}
\usepackage{geometry}
\usepackage{subfigure}
\usepackage{url}

\newcommand{\bd}{\begin{displaymath}}
\newcommand{\ed}{\end{displaymath}}

\newcommand{\be}{\begin{equation}}
\newcommand{\ee}{\end{equation}}
\newcommand{\br}{\begin{eqnarray}}
\newcommand{\bea}{\begin{eqnarray}}
\newcommand{\eea}{\end{eqnarray}}
\newcommand{\er}{\end{eqnarray}}
\newcommand{\ba}{\begin{array}}
\newcommand{\ea}{\end{array}}
\newcommand{\bi}{\begin{itemize}}
\newcommand{\ei}{\end{itemize}}
\newcommand{\bn}{\begin{enumerate}}
\newcommand{\en}{\end{enumerate}}
\newcommand{\bc}{\begin{center}}
\newcommand{\ec}{\end{center}}

\newcommand{\bs}{\begin{small}}
\newcommand{\es}{\end{small}}
\newcommand{\bfo}{\begin{footnotesize}}
\newcommand{\efo}{\end{footnotesize}}

 \def\({\left(}
 \def\){\right)}
 \def\[{\left[}
 \def\]{\right]}

\def\mHu{{m_{h_u}}}
 \def\mHd{{m_{h_d}}}
 \def\mHu2{{m_{h_u}^2}}
 \def\mHd2{{m_{h_d}^2}}

\def\haas{$H\to\gamma\gamma\;$}
\def\tqhs{$q\, b \to t \,q' H\;$}

\def\braa{$BR_{\gamma \gamma}$}

\def\tqhpp{$p\, p \to t \,q\, H$}
\def\tqaapp{$p\, p \to t \,q\, H\to t \,q\, \gamma \gamma$}
\def\tqhspp{$p\, p \to t \,q\, H\;$}
\def\tqaaspp{$p\, p \to t \,q\, H\to t \,q\, \gamma \gamma\;$}

\def\q2 {q^2}

\def\bt{\begin{table}}
\def\et{\end{table}}

\catcode`@=11 
\def \gsim{\mathrel{\mathpalette\@versim>}}
\def \lsim{\mathrel{\mathpalette\@versim<}}
\def \@versim#1#2{\lower0.4ex\vbox{\baselineskip\z@skip\lineskip\z@skip
     \lineskiplimit\z@\ialign{$\m@th#1\hfil##\hfil$%
     \crcr#2\crcr\sim\crcr}}}
\catcode`@=12 

\def\gappeq{\mathrel{\rlap {\raise.5ex\hbox{$>$}}
{\lower.5ex\hbox{$\sim$}}}}

\def\lappeq{\mathrel{\rlap{\raise.5ex\hbox{$<$}}
{\lower.5ex\hbox{$\sim$}}}}

\begin{document}
\pagestyle{empty}
\begin{center}
{\LARGE {\bf 
Direct constraints on the top-Higgs coupling
\vspace{0.1cm}
from the 8 TeV LHC data\\
}} 
\vspace*{1.5cm}
{\large
 {\bf Sanjoy Biswas$^{a}$, Emidio Gabrielli$^{{b,c,}}$\footnote{
On leave of absence from Dipart. di Fisica, Universit\`a di 
Trieste, Strada Costiera 11, I-34151 Trieste, Italy.}, Fabrizio Margaroli$^{d}$, and 
\\
\vspace{0.3 cm}
Barbara Mele$^{a}$}}
\vspace{0.3cm}

{\it
 (a) INFN, Sezione di Roma, \\ c/o Dipart. di Fisica, Universit\`a di Roma ``La Sapienza", \\ P.le Aldo Moro 2, I-00185 Rome, Italy}  \\[1mm]
{\it
(b) NICPB, Ravala 10, Tallinn 10143, Estonia}  \\[1mm]
{\it
 (c) INFN, Sezione di Trieste, Via Valerio 2, I-34127 Trieste, Italy}   \\[1mm]
{\it
 (d) Dipart. di Fisica, Universit\`a di Roma ``La Sapienza", \\
and INFN Sezione di Roma, \\P.le Aldo Moro 2, I-00185 Rome, Italy}

\vspace*{2cm}{\bf Abstract} \\
\end{center}
\vspace*{5mm}
The LHC experiments have analyzed the 7 and 8 TeV LHC data in the main Higgs production and decay modes. Current analyses only loosely 
 constrain an anomalous top-Higgs coupling in a direct way. In order to  strongly constrain this  coupling, the Higgs-top associated production 
is reanalyzed. Thanks to the strong destructive interference in the $t$-channel for standard model couplings, this process can be very sensitive to 
both the magnitude and the sign of a non-standard top-Higgs coupling. We project the sensitivity to anomalous couplings to the integrated luminosity 
of 50\,fb$^{-1}$, corresponding to the data collected by the ATLAS and CMS experiments in 7  and 8 TeV collisions,  as of 2012. We show that the combination 
of diphoton and multi-lepton signatures, originating from different combinations of the top and Higgs decay modes, can be a potential probe to constrain a 
large portion of the negative top-Higgs coupling space presently allowed by the ATLAS and CMS global fits.
\vfill\eject

\setcounter{page}{1}
\pagestyle{plain}

\section{Introduction}

The discovery of a Higgs-like resonance at the LHC \cite{Aad:2012tfa, Chatrchyan:2012ufa} has started up a new phase in the experimental exploration of the electroweak symmetry-breaking (EWSB) mechanism of the standard model (SM). The observed resonance is, within present experimental errors, well compatible with the minimal structure of a Higgs sector. Nevertheless, the determination of different properties of the new particle with increasing precision in the next years of the LHC running is expected to be a powerful mean to explore what could be beyond the SM description of fundamental interactions. In particular, on the one hand, deviations in the new-particle couplings to the electroweak vector bosons $V=W,Z$ would require further degrees of freedom to keep  the $VV$ scattering unitary. On the other hand, anomalies in the Yukawa couplings to matter fields could possibly point to a non-standard  mechanism for the generation of quark and charged-lepton masses.

With the present moderate statistics collected at 7 and 8 TeV  ($\sim$ 25 fb$^{-1}$ per experiment), where the main  analyses are based on the $H\to
\gamma\gamma,ZZ^*,WW^*$ signatures, global fits for the Higgs-like particle couplings  are made under simplified assumptions \cite{LHC:2012nn,atlas-11,cms-11}.
Similar strategies are followed
 in Higgs analysis at Tevatron, based on $\sim$ 10 fb$^{-1}$ in $p\bar p$ collisions at 1.96 TeV \cite{Aaltonen:2013kxa}.
In particular, one can assume a {\it universal}  scale factor $C_f$ (either positive or negative) for the Higgs  Yukawa couplings to all fermion species $f$,
\be
Y_f = C_f \; Y_f^{SM}\, ,
\ee
(where {$Y_f^{SM}=m_f/{v}$} is the $SM$ Yukawa coupling and $v=\langle H \rangle$ is the Higgs vacuum expectation value)\footnote{Note that in the SM,  the spontaneous EWSB does not fix  the overall sign of the Yukawa couplings, which can be rotated away by a chiral transformation of fermion fields.  What is observable is the relative sign of the Yukawa coupling and the corresponding fermion mass term in the Lagrangian (predicted as positive in the SM) that is the quantity $sign[Y_f\, m_f]$.
In the following, we set the sign of the fermion mass $m_f$ according to  the standard notation of the free fermion Lagrangian, and refer to the sign of the Yukawa coupling with respect to the $HWW$ coupling.}, 
and a {\it universal} (according to custodial symmetry) coupling scale factor $C_V$ to $W$ and $Z$ bosons, 
\be
 C_V =\; g_{HWW}/g_{HWW} ^{SM}=g_{HZZ}/g_{HZZ} ^{SM}\, .
\ee
The ATLAS and CMS two-dimensional fits are both compatible within $2\,\sigma$ with a SM coupling setup $C_V=C_f=1$ \cite{atlas-11,cms-11}. They are likewise compatible with a 
non-SM setup $C_V\simeq-C_f\simeq1$. The latter  two solutions would be degenerate if it were not for the presence in the production rates of   linear terms in $C_V$ and $C_f$ mainly associated to the $H\to \gamma \gamma$ decay width. The latter arise from the (destructive) interference of the top and $W$ loop amplitudes contributing to the decay, and tend to move the second minimum of the  $(C_V,C_f)$ likelihood function away from the $C_V\simeq-C_f\simeq 1$ region.
It has initially been argued  that resolving experimentally this ambiguity by improving the accuracy of the ATLAS and CMS fits will require a larger statistical sample than the present one  \cite{Espinosa:2012ir,Azatov:2012rd}. More recently, global fits of Higgs couplings tend to disfavor the $C_V\simeq-C_f\simeq 1$ solution 
\cite{Giardino:2013bma}. Nevertheless, the best-fit parameter values seem presently still quite sensitive to statistical fluctuations.

Alternative strategies aimed at excluding more directly the non-standard setup 
$C_f\simeq-C_V$ have been suggested. In \cite{Biswas:2012bd,Farina:2012xp}, the associated production of a single top quark and a Higgs boson \cite{Stirling:1992fx}-\cite{Bordes:1992jy} was proposed as a direct probe of the relative sign of the $Ht\bar{t}$ and $HWW$ couplings at the LHC. In particular, the 
$t$ channel 
\be
q\, b \to t \,q' H
\ee
(where the jet originating from the final light quark tends to be at large rapidities) 
is remarkably sensitive to a change in the $C_t$ relative sign, because of the large destructive interference (proportional to  $\sim Y_t\,m_t$) of the two dominant amplitudes of the process in the SM 
(Figure 1) \cite{Tait:2000sh}-\cite{Barger:2009ky}. 
\begin{figure}[t]
\begin{center}
\includegraphics[width=0.5\textwidth]{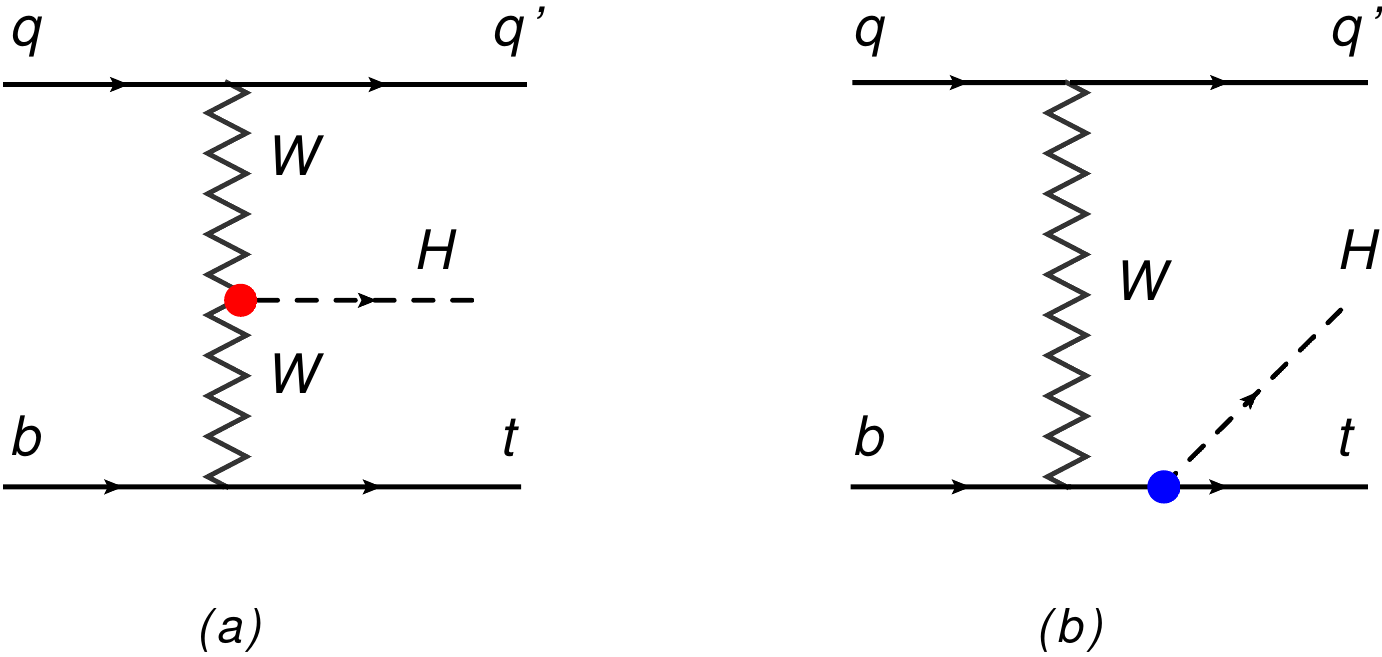}
\vspace{-0.3cm}
\caption{\small \it Feynman diagrams for the  single-top plus Higgs associated production 
in the $t$-channel \tqhs at  hadron colliders.  Higgs radiation by the initial $b$-quark line is neglected. }
\label{fig:effvert}
\end{center}
\end{figure}
 By flipping the
$Y_t$ sign, 
the total production cross section for \tqhspp gets enhanced by a factor $\sim13$ over the SM value of $\sim15$ (72) fb at 8 (14) TeV (summing over $t$ and $\bar t$ production)\footnote{Note that NLO QCD corrections increase the SM cross sections by about 15\% \cite{Farina:2012xp,Campbell:2013yla}.}, 
leading to about 0.21 (0.89) pb, for $C_t\simeq -1$ \cite{Biswas:2012bd}. As a consequence, while a study of the \tqhspp process in the SM
will need quite large integrated luminosities at 14 TeV, the 
production rates for $C_t\simeq-C_V\simeq-1$ can be large enough to start a direct exploration of  the reversed $Y_t$ setup with the integrated luminosity already collected at 7 and 8 TeV.

Note that a change in the  Yukawa sign and/or absolute value with respect to its SM value would 
signal an origin of the fermion masses different from spontaneous EWSB, and  spoil the unitarity  and renormalizability of the theory \cite{Appelquist:1987cf2}. 
In particular, it was shown in \cite{Farina:2012xp} that a change in the relative sign of the Higgs $t$ and $W$ couplings in the $Wb\to tH$ process (the relevant subprocess of the \tqhs 
scattering)
 leads to a violation of the perturbative unitarity at a scale $\Lambda\simeq10$ TeV. This  is well above the effective partonic c.m. energies of the $tH$ system 
involved in the \tqhs production at the LHC (which is at most $\sim 1$ TeV), and 
 implies that a perturbative calculation of the \tqhs cross section at the LHC collision energies can still be reliable.

In order to assess the LHC potential for detecting an enhancement in the \tqhs cross section, one can exploit different
$H$ and $t$-quark decay channels. As in the main Higgs search processes,
considering non-strongly interacting decay products (like in  $H\to \gamma\gamma,\;
\ell\ell\ell'\ell',\; \ell\ell' \nu \nu'$ and $t\to q q'  b$ or $t\to \ell\nu b$) leads to a reduction in  rates, which is more than compensated by the cleaner experimental signatures, that suffer from milder backgrounds. 
\\
All Higgs decay branching ratios (BR's) are quite affected by changes in the Yukawa sector.
While the \tqhs total cross section is  sensitive
to  the Yukawa sector  mostly through the $t$-quark coupling, the Higgs decay BR's are all  sensibly affected also by possible  changes in the magnitude of the Higgs couplings to lighter quarks and heavy leptons, notably $Y_b$, $Y_c$, $Y_{\tau}$\footnote{A change in the sign of the latter couplings would be nevertheless ineffective in this case.}.

For simplicity, and following the present experimental analysis \cite{LHC:2012nn,atlas-11,cms-11}, we will assume from now on either a {\it universal}  scale factor $C_f$ (varying in the range $-1.5\lappeq C_f\lappeq 1.5$) for 
all fermion species $f$, or a scale factor $C_t$ parametrizing the top Yukawa coupling, with $C_f=1$ for all lighter fermions\footnote{In both cases, 
loop-induced processes are assumed to scale according to the SM loop structure.}. 
As for the $HVV$ couplings, we will allow for a variation of a few
10\% around the SM value $C_V=1$.

In \cite{Biswas:2012bd}, a study of the potential of the \tqhs channel for revealing a non-standard Yukawa sign was made, exploiting the two-photon signature for the Higgs decay accompanied by hadronic top decays plus a forward jet
\be
q\; b \to \,t \,q'\, H \to  \, (b \, q q')\,q' \,\gamma \gamma\, .
\label{old}
\ee
A parton-level simulation, including the main irreducible backgrounds,  in the universal $C_f$ hypothesis, found that  at 14 TeV, for $-1.5\lappeq C_f\lappeq 0$,  an integrated luminosity of $60$ fb$^{-1}$
would give about 10 signal events versus less than 0.3 background events, with a corresponding signal significance at a 3-$\sigma$ level. At 8 TeV, with same integrated luminosity, the corresponding significance  drops to around $1.5 \,\sigma$.

In \cite{Farina:2012xp}, the potential of \tqhspp  was assessed through the hadronic
$H\to b\bar b$ decay, focussing on the quite challenging 3-$b$ and 4-$b$ signatures,  and semileptonic $t$ decay 
\be
q\, b\;  \to \, t \,q'\,H\to   \,( \ell\,\nu\, b)\,q' \,b\,\bar b \;,\;\;\;\;\;\;\;
q\,g \to \, t \,q'\,H\,b\to \, (\ell\,\nu\, b) \,q'  b\,\bar b \, b\;.
\ee
In the universal $C_f$  hypothesis, it turns out that 50 fb$^{-1}$ at 14 TeV  could be sufficient to exclude the negative $C_f$ range presently allowed by experimental fits, while at 8 TeV the same integrated luminosity would only partially exclude this possibility.  On the other hand,  after 
 switching to a non-universal $C_t$ scaling with  $C_b=C_c=C_{\tau}=1$, 
the reach  of the $H\to \bar b b$ channel is enhanced in the region  $|C_t| \lappeq 1$, because of  the corresponding relative increase in the $H\to \bar b b$ width.
 
 The aim of the present analysis is to extend our previous study of the \tqhs channel
 in \cite{Biswas:2012bd} in order to assess its potential for the exclusion of a 
 non-standard Yukawa sign with the statistics presently accumulated at 7 TeV and 8 TeV at the LHC.  We present a parton-level
 analysis including   the most robust signatures, corresponding to either photon pairs or
 more than one lepton in the final states, for which
 the rate suppression by the corresponding Higgs decay $BR$ is expected to be compensated by a better   signal-to-background ratio $S/B$. The latter signatures are 
 associated to the decays 
 \be
H\to \gamma\gamma \; ,\;\;\;\;
H\to WW^*\to \ell\nu \ell'\nu,\; \ell \nu\,qq'\, , \;\;\;\; H\to \tau \tau \to \ell\nu\nu \, \ell'\nu\nu,\; \ell\nu\nu + had(s)\, ,
\ee
which, in the $C_f\lappeq 0$ case, are expected to be statistically relevant in  the \tqhs production  with the present LHC data set.
In particular, on the one hand, we complement  the $H\to \gamma\gamma\;$ analysis 
for a hadronic top decay  [cf. Eq.~(\ref{old})]  carried out in \cite{Biswas:2012bd} with the 
one with a semi-leptonic top decay, requiring two photons plus a charged lepton in the final state
\be
q\; b \to \,t \,q'\, H \to  \, ( \ell\,\nu\, b)\,q' \,\gamma \gamma\, .
\label{new1}
\ee
On the other hand, we consider now also: 
{\it i)} the three-lepton signature
arising from the semi-leptonic top decay combined with the decays 
$H\to WW^*\to \ell\nu \ell'\nu,$ and $H\to \tau \tau \to \ell\nu\nu \, \ell'\nu\nu$, and 
{\it ii)} the two-same-sign-lepton signature 
arising from the semi-leptonic top plus  the decays  
$H\to WW^*\to \ell\nu \ell'\nu, \; \ell \nu\,qq'$,
and $H\to \tau \tau \to \ell\nu\nu \, \ell'\nu\nu,\; \ell\nu\nu + had(s)\,$,
 leading to the following final states
\be
q\, b\;  \to \, t \,q'\,H\to   \,( \ell\,\nu\, b)\,q' \,\ell \,\ell'+\nu's\;,\;\;\;\;\;\;\;
q\, b\;  \to \, t \,q'\,H\to   \,( \ell\,\nu\, b)\,q' \,\ell +\nu's+X_h\,,
\label{new2}
\ee
where  $X_h$  is made of further hadronic states either from the $W$ or the $\tau$ hadronic decays.
In the present analysis, we closely follow the approach in  \cite{Biswas:2012bd},
and  estimate the corresponding main {\it irreducible} (and {\it reducible} for the diphoton channel) backgrounds at  parton level, requiring the tagging of a $b$-jet  
in the final state (arising from the top decay), plus a forward light jet\footnote{Reducible backgrounds are expected to be subdominant for the multilepton signatures considered here.}.

We will show that, using the present LHC statistics, the  $H\to \gamma\gamma,WW^*,\tau\tau$ decays in  \tqhspp have a strong potential   to exclude in a direct way most of the  
$C_f\lappeq 0\,$ region  in the $(C_V,C_f)$  plane which is allowed  by the present ATLAS and CMS Higgs-coupling fits. As expected, the exclusion power of the photon-pair and 
multi-lepton signatures (excluding the $H\to \tau\tau$ component) turns out to be complementary 
to the $H\to b\bar b$ decay, the corresponding rates being enhanced when $|C_b|\lappeq 1$, where the $H\to b\bar b$ signal rates are depleted.

The plan of the paper is the following. In Section 2, we detail the $(C_V,C_f)$  dependence of the  \tqhspp  production cross sections, and the corresponding signal event rates for 
the $H\to \gamma\gamma,WW^*,\tau\tau$ decays. In Section 3, we describe the relevant SM backgrounds and selection cuts for the processes in  Eqs.(\ref{new1}) and (\ref{new2}), 
and recall the relevant results of our  previous study for  Eq.(\ref{old}). For
a data set of 50 fb$^{-1}$, we present for each signature the 
 event numbers (S) for signals in different channels versus $C_t=0,\pm 1$ (and $C_V=1$), the  background event numbers (B),
and the corresponding expected significance.
In Section 4, we combine our findings, and show the expected exclusion regions in the  $(C_V,C_f)$ plane, both in the universal Yukawa scaling assumption, and in  the  hypothesis of $C_t$ scaling with $C_{f\neq t}=1$. In Section 5, we draw our conclusions.

\section{Signal production rates versus $C_V$ and $C_f$}
In this section, we present the \tqhspp cross-section and $BR(H\to \gamma\gamma,WW^*,\tau\tau)$ dependence on the $C_f$ and $C_V$ scaling factors through the corresponding enhancement factors $R_i$ with respect to the SM predictions. 
In particular, we will detail the behavior of the quantities 
\be
R_{\sigma}=\frac{\sigma}{\sigma^{SM}}  \; ,\;\;\;\;\;\;\;\;\;\;\;\;
R_{BR_{\gamma\gamma,WW,\tau\tau}}=\frac{BR_{\gamma\gamma,WW,\tau\tau}}{( BR_{\gamma\gamma,WW,\tau\tau})^{SM}} \;\;.
\label{ratios}
\ee
By combining the latter, we get  the signal strength for the corresponding decay channel
\be
R_{\sigma\cdot BR_{\gamma\gamma,WW,\tau\tau}}=\frac{\sigma\cdot BR_{\gamma\gamma,WW,\tau\tau}}{(\sigma\cdot BR_{\gamma\gamma,WW,\tau\tau})^{SM}} \;\;,
\label{ratio-yield}
\ee
that will primarily affect the signal event number dependence on the 
$C_f$ and $C_V$ scaling factors. A further moderate coupling dependence will arise from the efficiencies of actual kinematic cuts in the final event selection (cf. Section 3).
  
We will make two different assumptions for the Yukawa coupling variation :
\begin{itemize}
\item
{\it Universal Yukawa rescaling}, which  assumes  just one free parameter  $C_f=C_t$ describing all Yukawa couplings both in production and decay amplitudes. Varying
the top-quark coupling $C_t$ will then largely affect not only the production cross section, but also the Higgs branching ratios 
$BR_{\gamma\gamma}$, $BR_{WW}$ and  $BR_{\tau\tau}$. The main effect of the $C_f$ variation on $BR_{WW}$ will arise from the Higgs total width dependence on the Higgs couplings to the $b$ quark and lighter fermions. On the other hand, in $BR_{\gamma\gamma}$  there will be a further non-trivial dependence on  $C_t$ caused by the top-quark loop that in the SM interferes destructively 
with the $W$ loop in the $H\to \gamma\gamma$ amplitude.
\item
{\it  Free  $C_t$ and $C_{f\neq t}=1$}, with 
   $C_t$ affecting mainly the   production  cross sections and   the $H\to \gamma\gamma$ width, while  leaving almost unaltered  $BR_{WW}$ and  $BR_{\tau\tau}$.
\end{itemize}

In the following, all  numerical cross sections will refer to the hadronic $p\,p$ collisions (even when the {\it partonic} initial state is shown). The production rates at leading order have been computed via the MADGRAPH5 (v1.3.33) software package \cite{Alwall:2011uj}, and the CTEQ6L1 
 parton distribution 
functions \cite{CTEQ6L1}. Both the  factorization and renormalization scales are set at the value  $Q=\frac{1}{2}(m_H+m_t)$ for the  \tqhspp 
signal, where $m_t$ is the top-quark mass. The other relevant parameters entering our computation  are set as follows
\cite{Aad:2012tfa,Chatrchyan:2012ufa,Aaltonen:2012ra,Beringer:1900zz}:
$m_H=125 {\rm ~GeV},\;~m_t=173.2 {\rm ~GeV}, \;
M_Z=91.188 {\rm ~GeV},\;~M_W=80.419 {\rm ~GeV}, \;
m_b=4.7 {\rm ~GeV},~{\rm and} \;~\alpha_S(M_Z)=0.118\;.$

The $BR_{\gamma\gamma}$, $BR_{WW}$ and  $BR_{\tau\tau}$ values versus the 
$C_f$ and $C_V$ scaling factors have been obtained through an original code we have built up, containing all the leading corrections (including  off-shellness effects for both $W$'s in the decay $H\to W^*W^*$)\footnote{We checked that its results for the SM case match the HDECAY~\cite{Djouadi:1997yw} predictions within a few per-cent.}.

\begin{figure}[thp]
\begin{center}
\vspace*{-4cm}
\centerline{
\hspace{-0.75cm}
\includegraphics[width=0.84\textwidth]{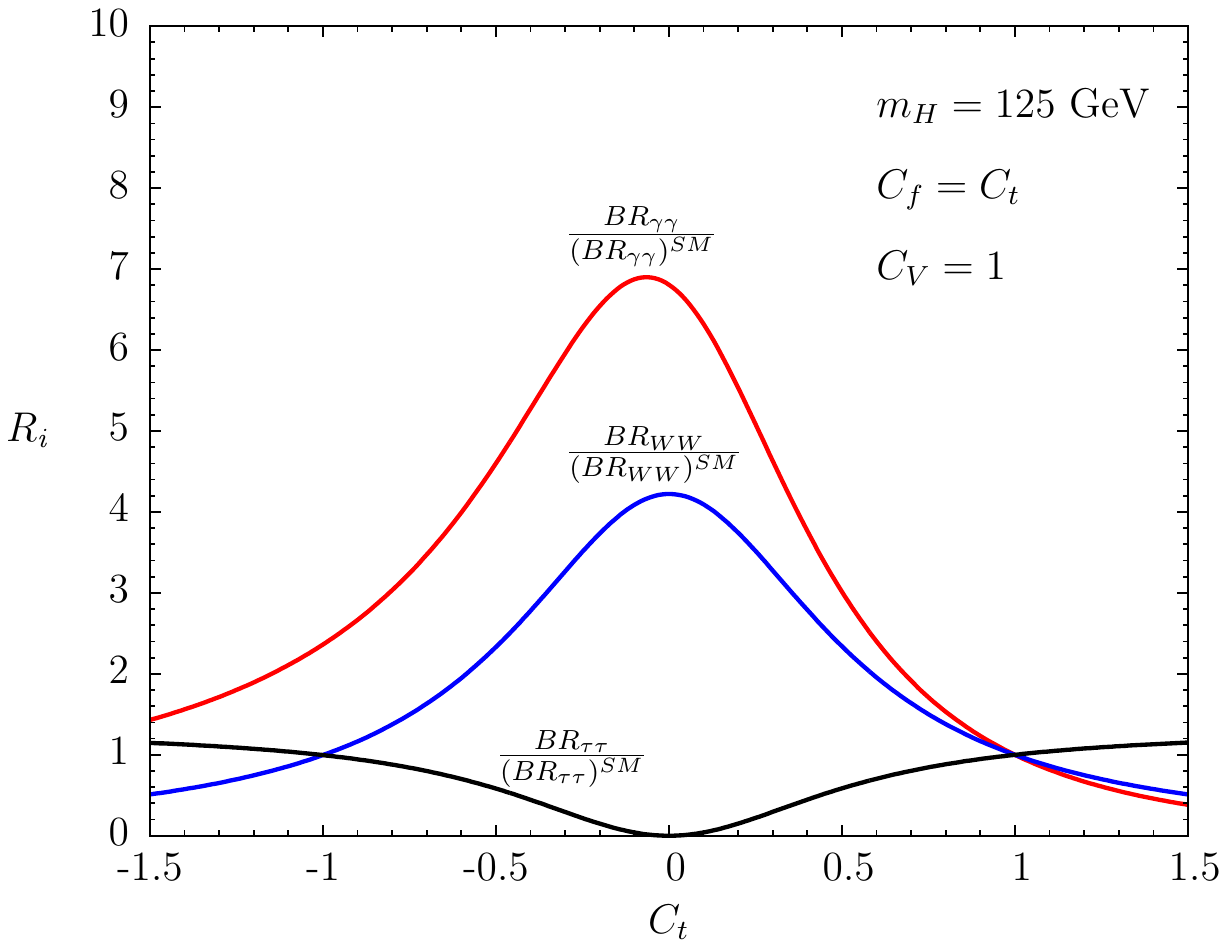}
\hspace{-6.1cm}
\includegraphics[width=0.84\textwidth]{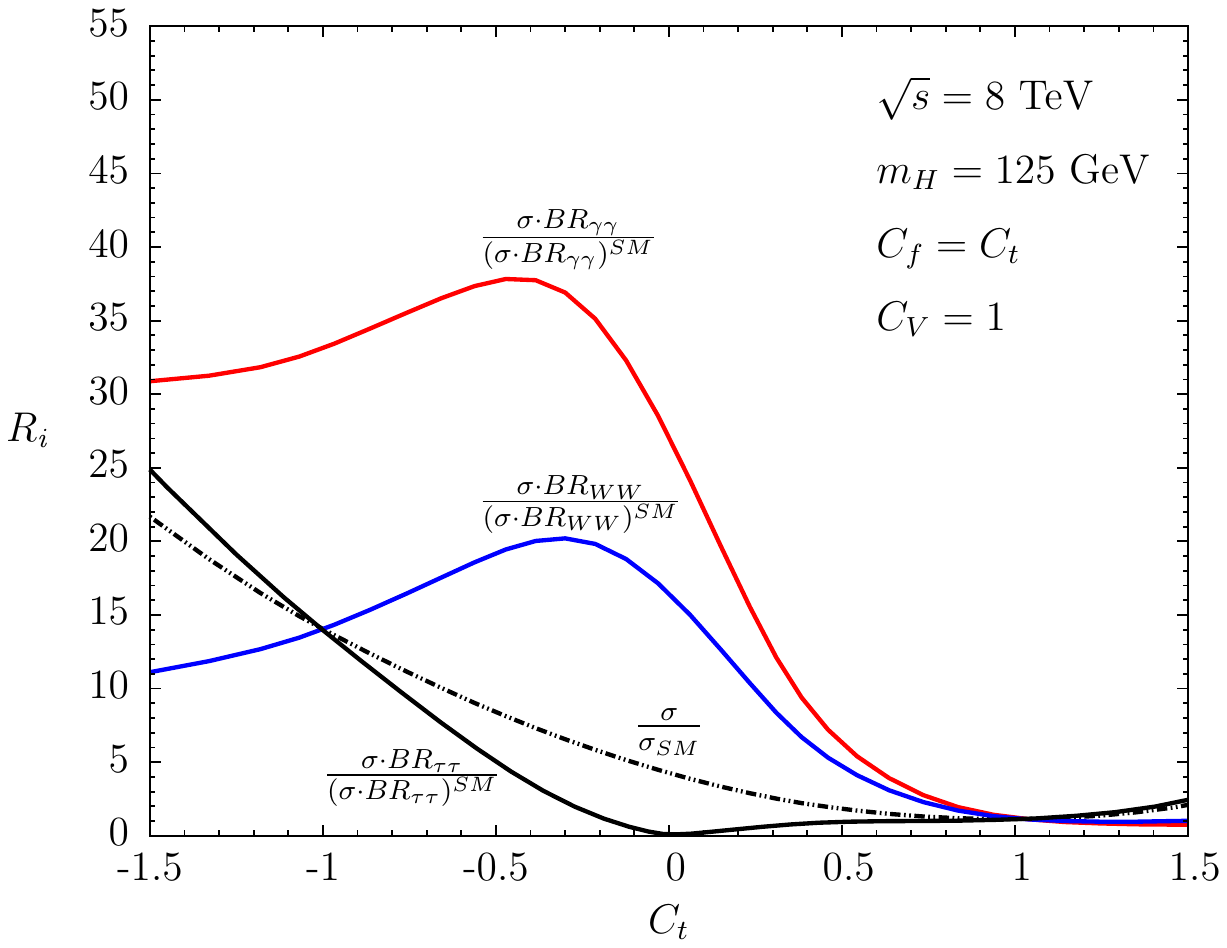}
}
\vskip 1pt
\vspace{-12cm}
\centerline{
\hspace{-0.75cm}
\includegraphics[width=0.84\textwidth]{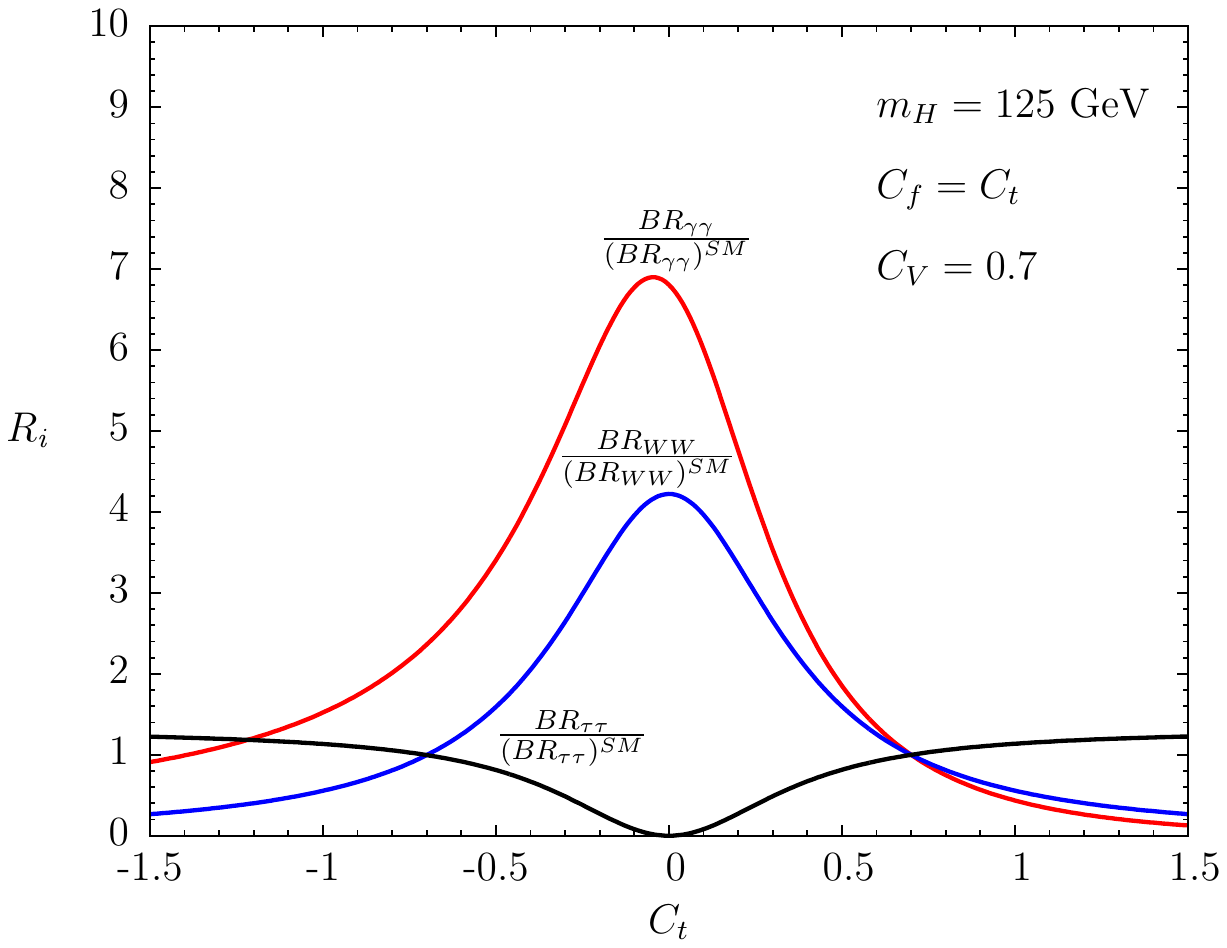}
\hspace{-6.1cm}
\includegraphics[width=0.84\textwidth]{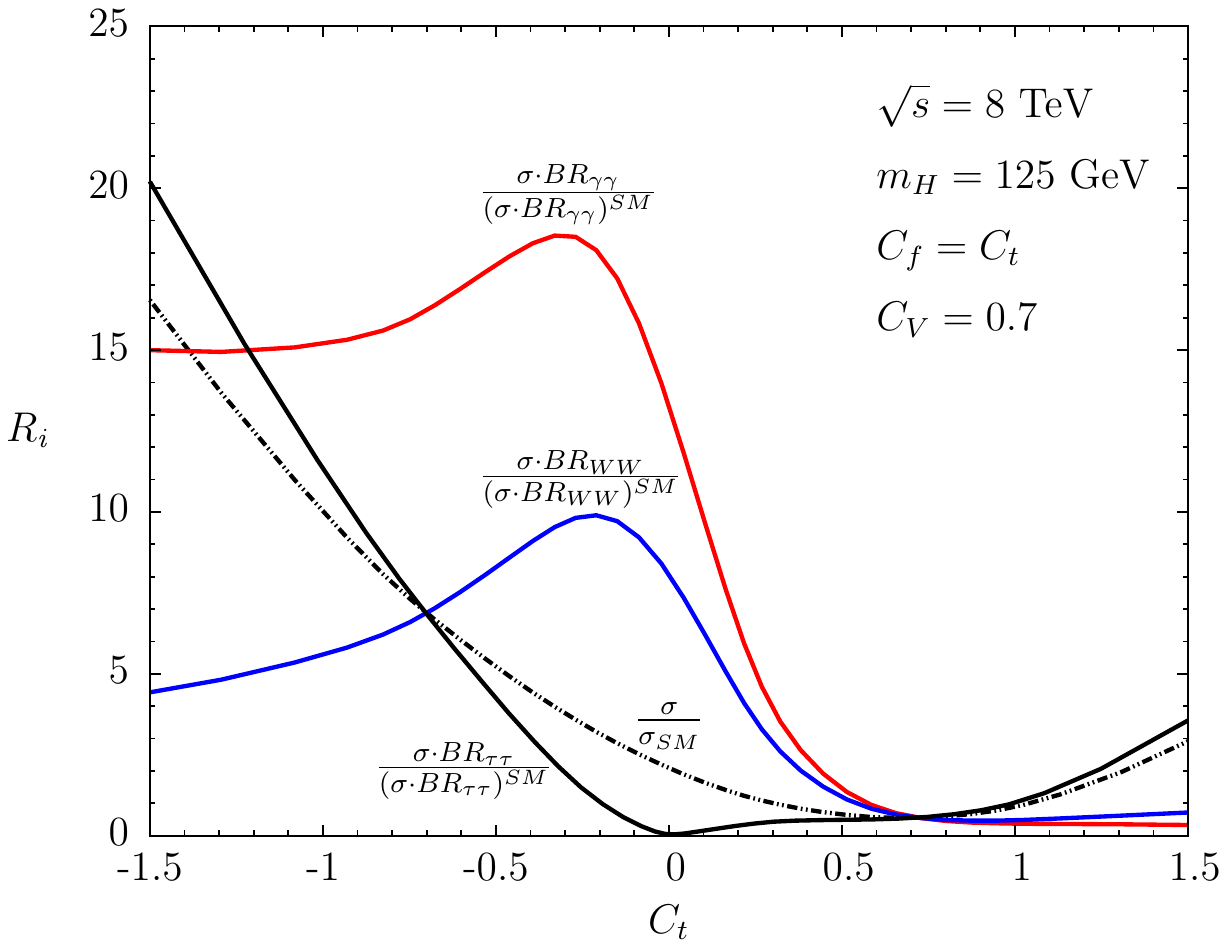}
}
\vskip 1pt
\vspace{-12cm}
\centerline{
\hspace{-0.75cm}
\includegraphics[width=0.84\textwidth]{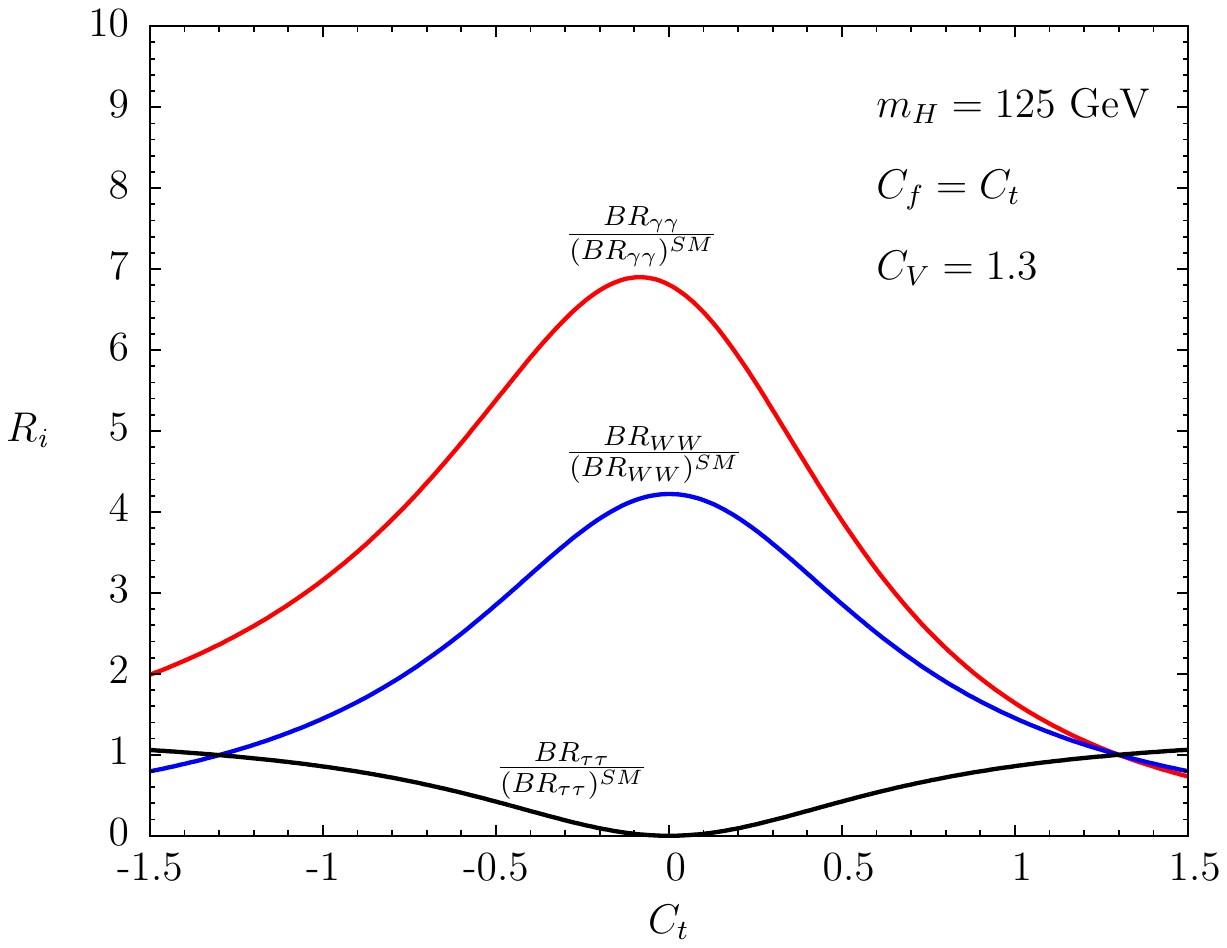}
\hspace{-6.1cm}
\includegraphics[width=0.84\textwidth]{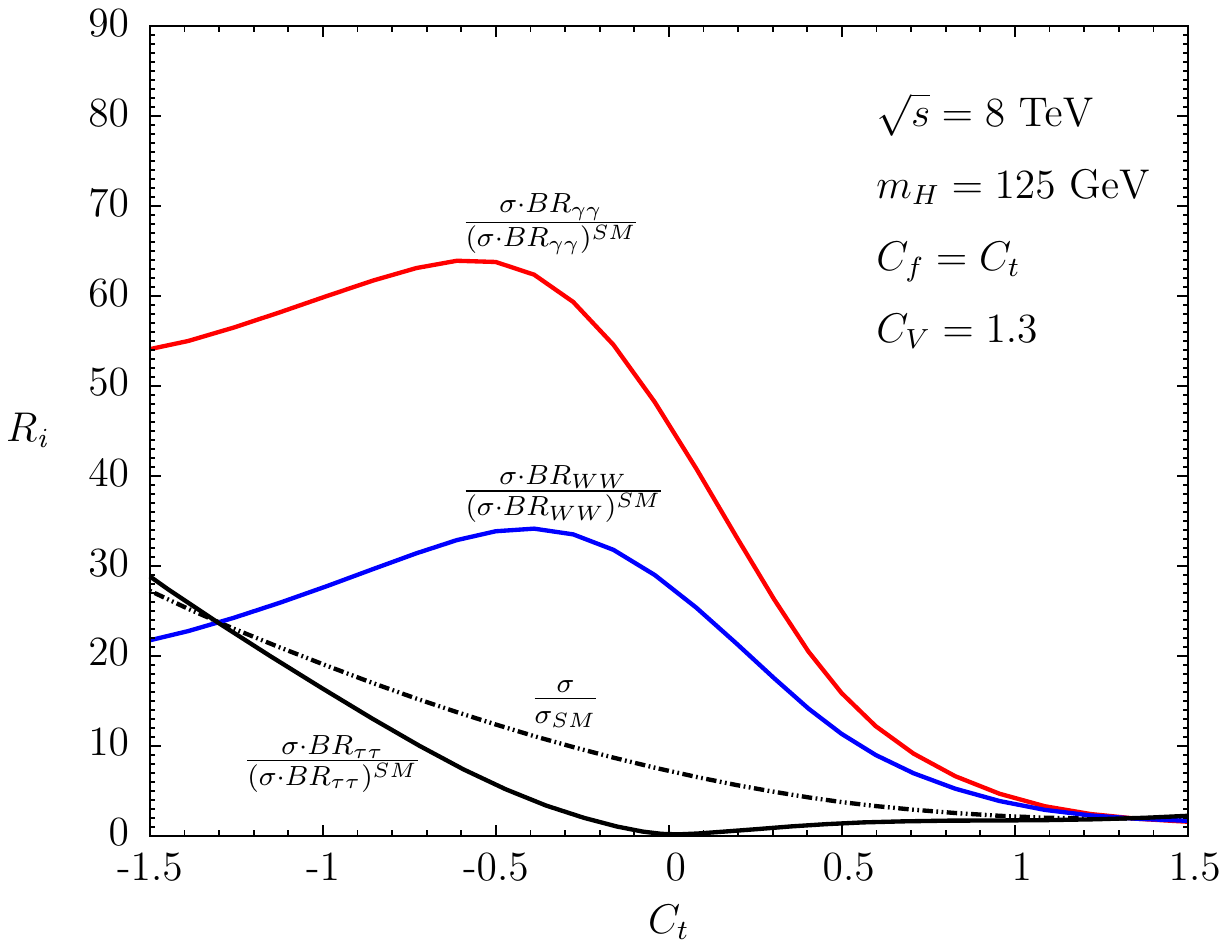}
}
\vspace{-8.5cm}
\caption{\small Enhancement factors $R_i$ versus $C_t$, normalized to
the SM, for the branching ratios 
$BR_{\gamma\gamma}$, $BR_{WW}$, and  $BR_{\tau\tau}$ (left), 
and the corresponding signal strengths  as 
defined in Eq.~(\ref{ratio-yield}) (right),  
 for  $\sqrt{S}= 8$ TeV, and for 
the universal scaling scenario  $C_{f}=C_t$.
Dashed lines in the right plots correspond to the $pp \to t q H$ 
total cross section $\sigma$ normalized to the SM one.  
Plots from top to bottom correspond to $C_V=1,0.7,1.3$, respectively.
}
\label{fig:ratio-A}
\end{center}
\end{figure}

\begin{figure}[thp]
\begin{center}
\vspace*{-4cm}
\centerline{
\hspace{-0.75cm}
\includegraphics[width=0.84\textwidth]{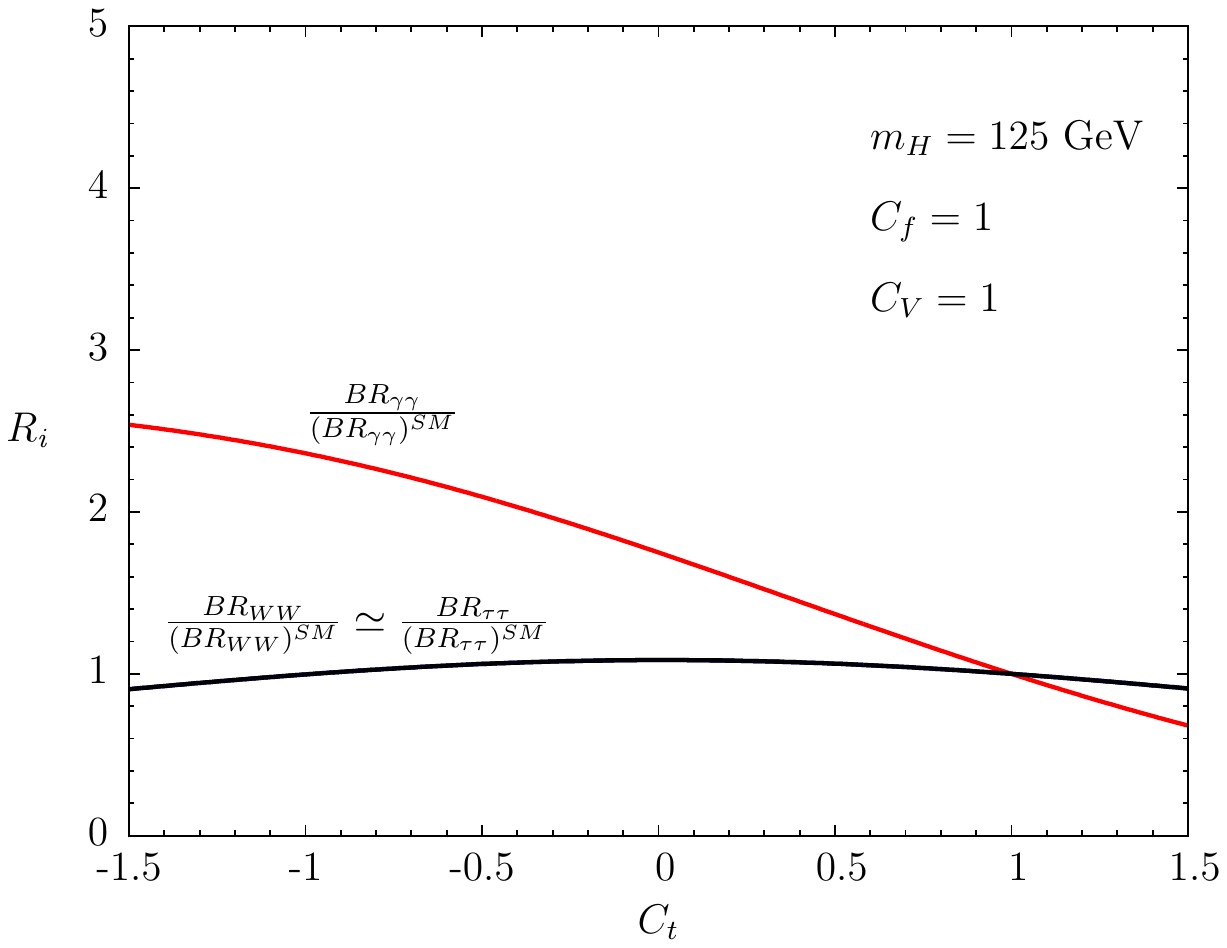}
\hspace{-6.1cm}
\includegraphics[width=0.84\textwidth]{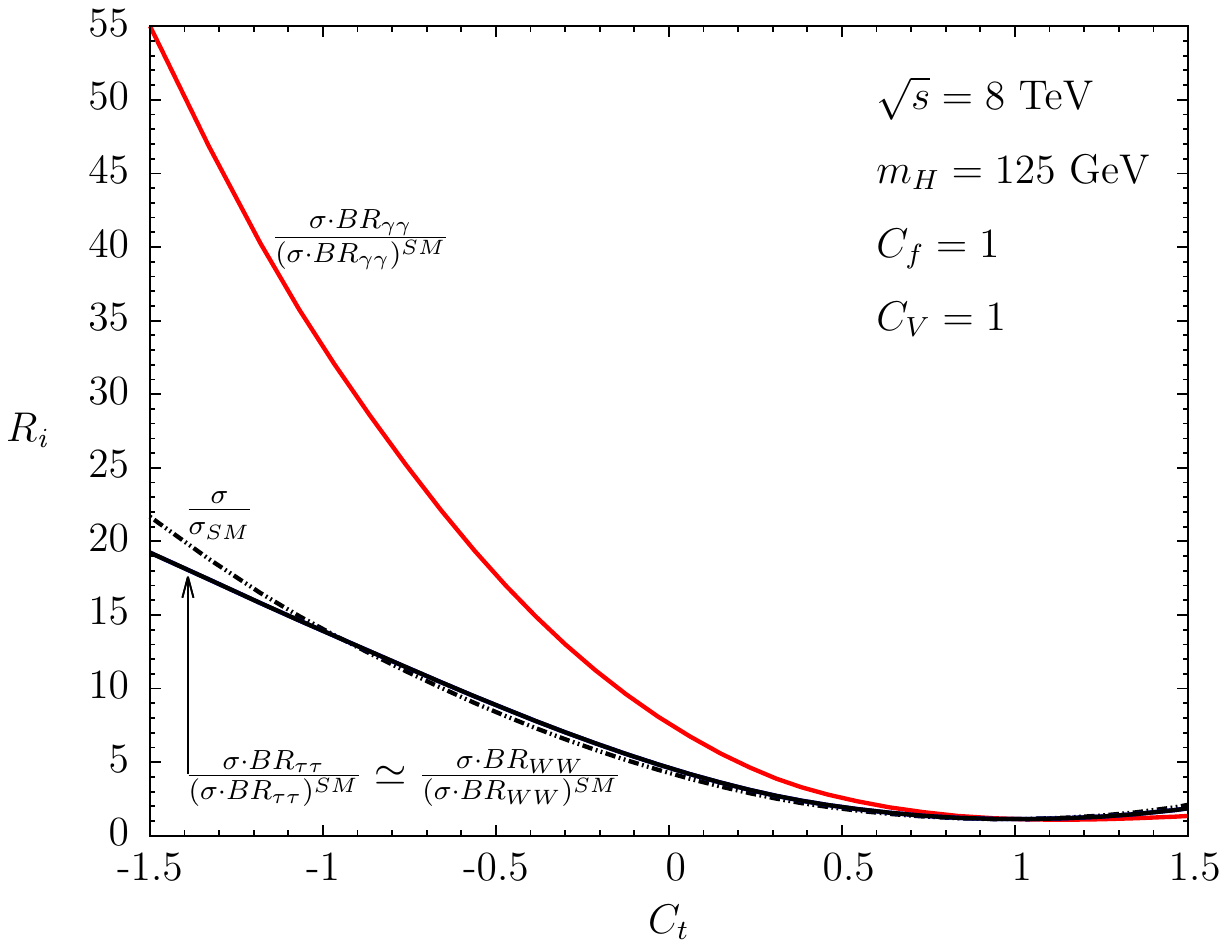}
}
\vskip 1pt
\vspace{-12cm}
\centerline{
\hspace{-0.75cm}
\includegraphics[width=0.84\textwidth]{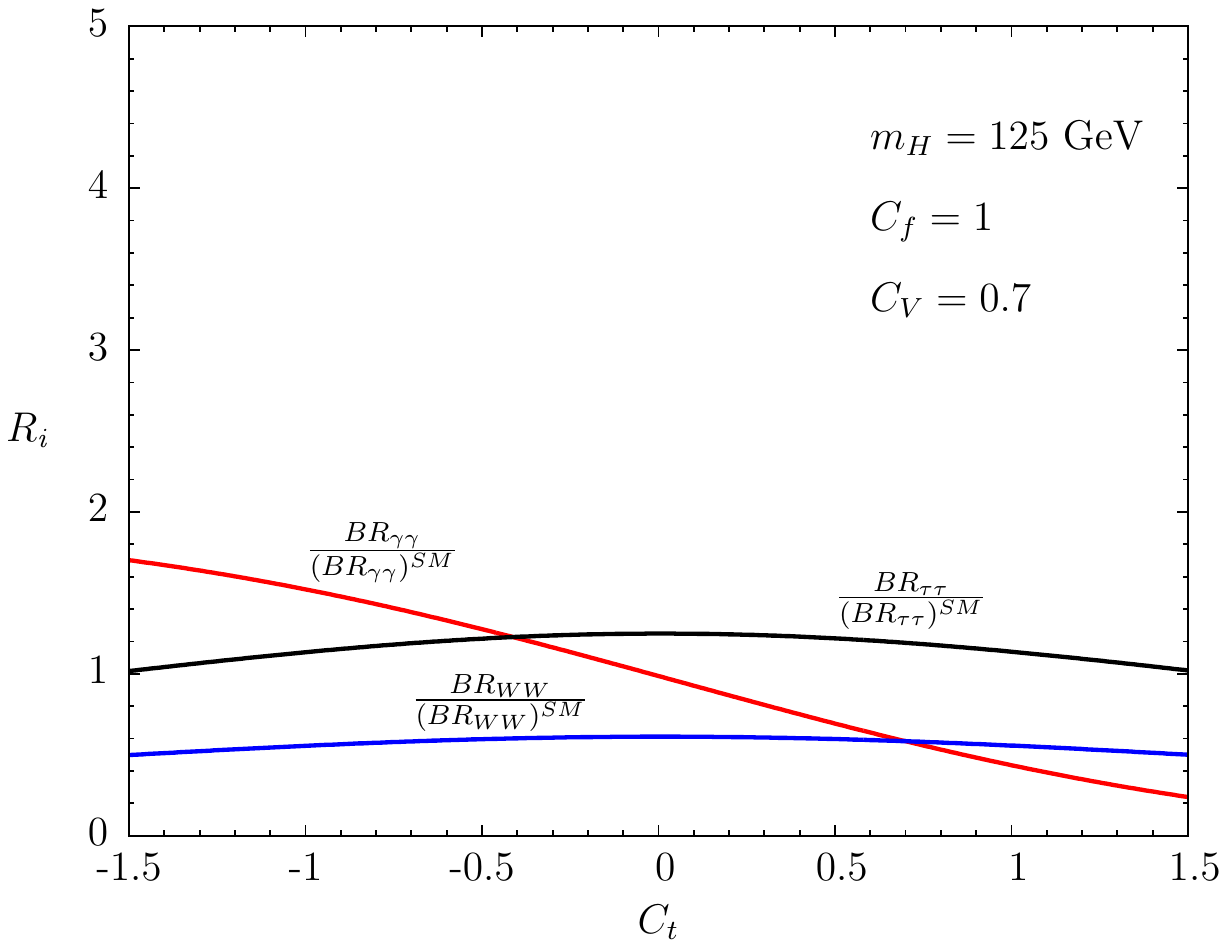}
\hspace{-6.1cm}
\includegraphics[width=0.84\textwidth]{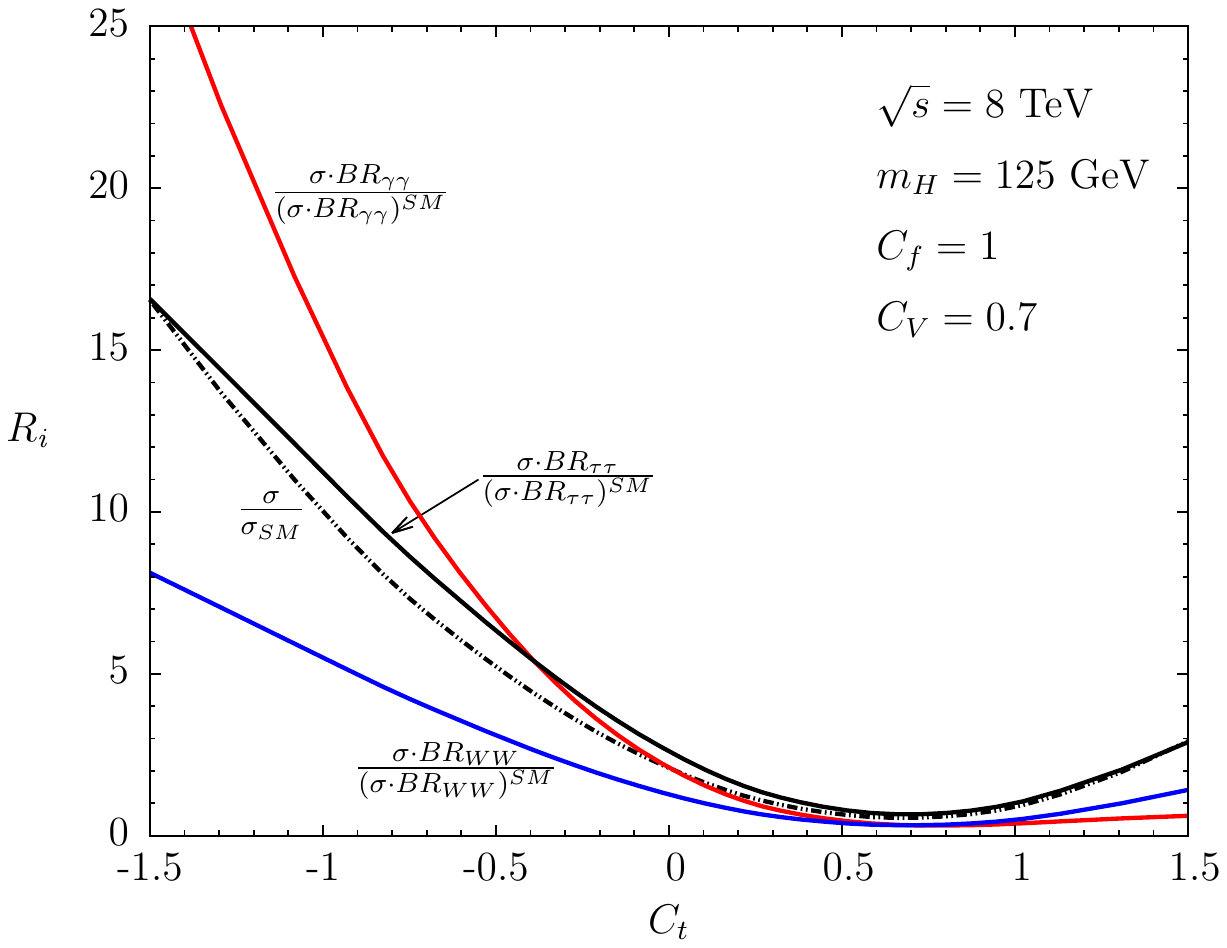}
}
\vskip 1pt
\vspace{-12cm}
\centerline{
\hspace{-0.75cm}
\includegraphics[width=0.84\textwidth]{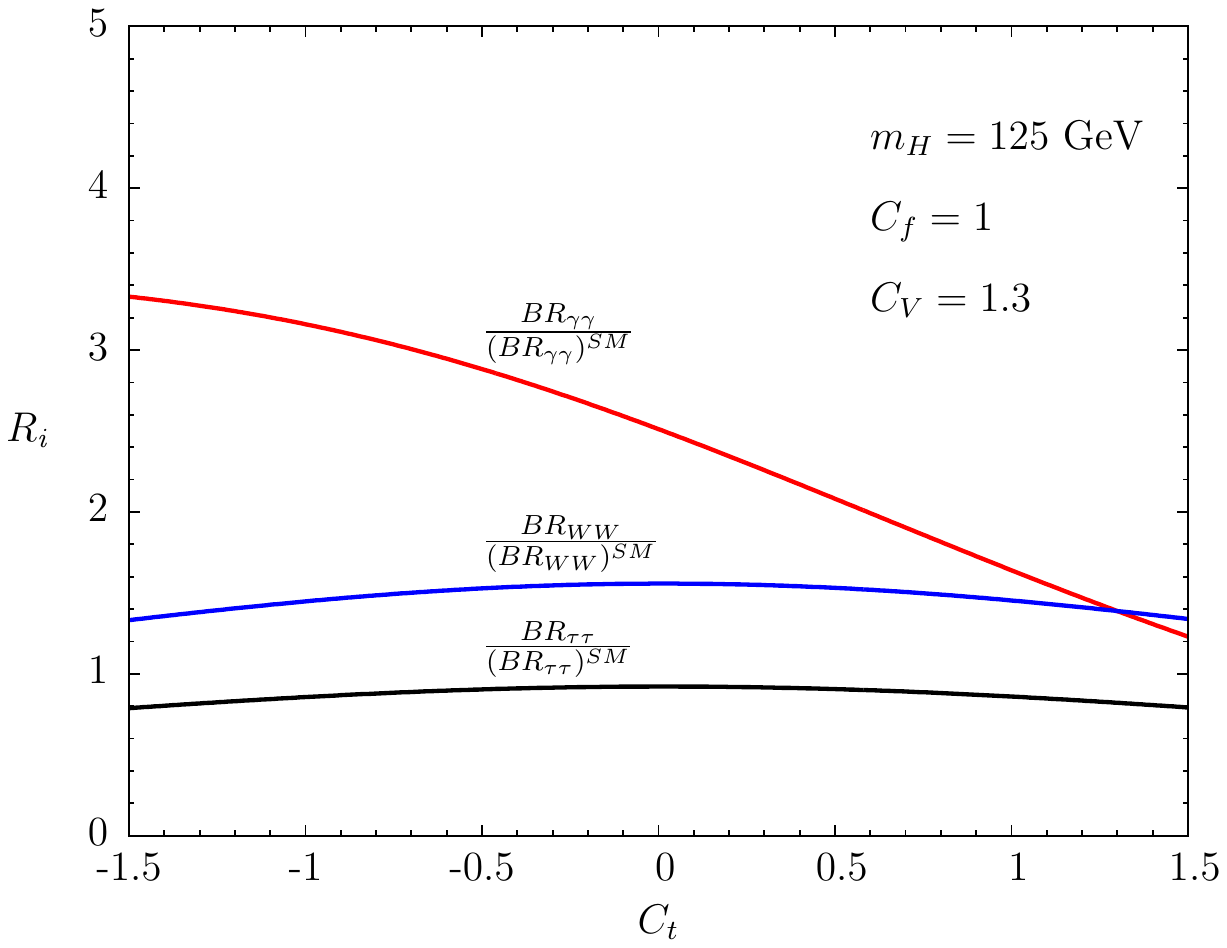}
\hspace{-6.1cm}
\includegraphics[width=0.84\textwidth]{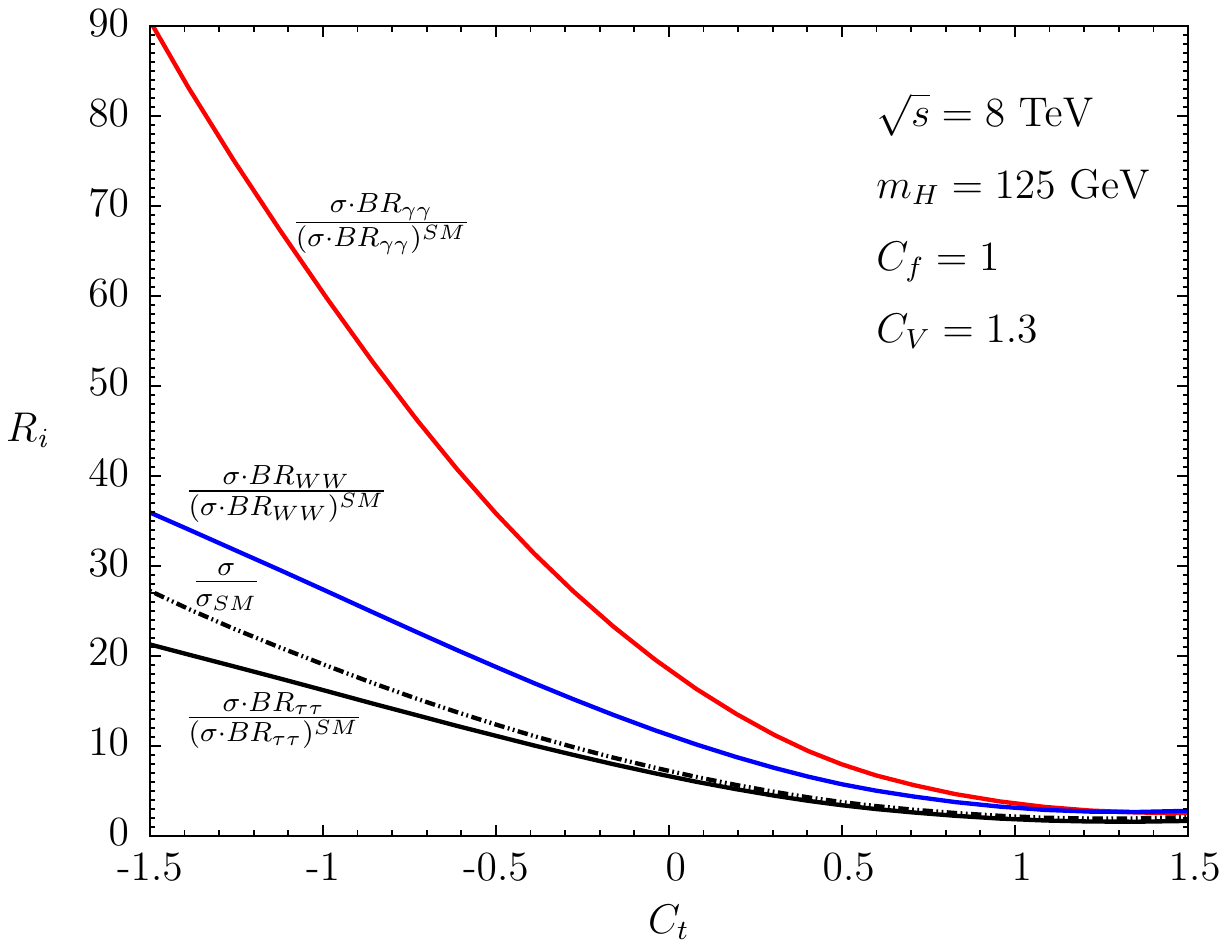}
}
\vspace{-8.5cm}
\caption{\small Same as in Figure \ref{fig:ratio-A}, but for 
the $C_t$ scaling scenario  with $C_{f\neq t}=1$.
}
\label{fig:ratio-B}
\end{center}
\end{figure}

For reference, the  $SM$  \tqhspp cross section (summing up contributions from the two charge-conjugated channels)\footnote{The contribution to the \tqhspp cross section of the amplitude where the  Higgs is radiated by the initial $b$-quark line is small (at the per-mil level in the $C_t$ range relevant here),  and will be neglected in the present analysis.} is 
\bea
\sigma(q\, b \to t \,q' H)^{SM}&\simeq& 15.2\; {\rm fb\;\;\;\;\;\;\;\;\;\; at} \;\;\;\sqrt s = 8\; {\rm TeV}\; ,
\eea
and the $SM$ Higgs branching ratios $BR_{\gamma\gamma}$, $BR_{WW}$ and  $BR_{\tau\tau}$, that are relevant in the following,  are 
\bea
BR^{SM}_{\gamma \gamma}\simeq 2.30\cdot 10^{-3} \, , \;\;\;\;\;\;\;\;\;
BR^{SM}_{WW}\simeq 0.209 \, , \;\;\;\;\;\;\;\;\;BR^{SM}_{\tau\tau}\simeq 6.55\cdot 10^{-2}\;.
\eea
Note that, for $C_V=1$, the total Higgs width $\Gamma_H$
drops,
 for vanishing $C_t$, 
from its $SM$ value  $\Gamma_H(C_t=\pm 1)\simeq 4.04$ MeV    to 
$\Gamma_H(C_t=0)\simeq 0.955$ MeV, assuming a universal scaling in $C_f$.

In Figure~\ref{fig:ratio-A} we show (for the three values $C_V=1,0.7,1.3$, from top to bottom)  the dependence on $C_t$ of the enhancement factors $R_i$ , as defined in Eq.~(\ref{ratios}), for 
$BR_{\gamma\gamma}$, $BR_{WW}$, $BR_{\tau\tau}$ (left), and the corresponding  signal strengths, as defined in Eq.~(\ref{ratio-yield}),  at $\sqrt s=$ 8 TeV (right). 
The red, blu, black lines refer to the \haas, 
$WW^*$,  $\tau\tau$ channels respectively. Plots in 
Figure~\ref{fig:ratio-A}  correspond to the {\it universal scaling} 
scenario $C_{f}=C_t$.
In Figure~\ref{fig:ratio-B}, we show the same plots
as in Figure~\ref{fig:ratio-A}, but for the $C_t$ scaling scenario with $C_{f\neq t}=1$.
In the right plots of Figures~\ref{fig:ratio-A} and \ref{fig:ratio-B}, we also show the corresponding enhancement for the basic \tqhspp total cross section 
(dashed-dot lines) normalized to the SM, versus $C_t$. 

Throughout the paper, we focus on the range  
\be
-1.5<C_t<1.5\; ,
\ee
where the $C_t$ dependence is more critical, and  the most favored regions of the LHC fits  lie~\cite{atlas-11,cms-11}. In the universal scaling scenario (cf. Figures~\ref{fig:ratio-A}, left), 
the $R_{BR_{\gamma\gamma}}$ and $R_{BR_{WW}}$ 
behavior  is clearly driven by the $\Gamma_H$ reduction for $|C_f|\to 0$,
with  asymmetric corrections due to interference effects enhancing 
$R_{BR_{\gamma\gamma}}$ for negative $C_t$. 
\\
In the $C_{f\neq t}=1$ scenario (cf. Figures~\ref{fig:ratio-B}, left), 
$R_{BR_{\gamma\gamma}}$ and $R_{BR_{WW}}$ are definitely less sensitive to
$C_t$, with the dominant effect deriving from the positive interference effects in 
$R_{BR_{\gamma\gamma}}$ for $C_t<0$.
\\
An even more dramatic enhancement for negative $C_t$ is  present in the \tqhspp production cross section. When combined with the $R_{BR_{\gamma\gamma}}$ and $R_{BR_{WW}}$ behavior for universal $C_f$ scaling (cf. Figures~\ref{fig:ratio-A}, right),
the latter tends to level up the dependence on $C_t$ of the corresponding signal strengths
in the negative $C_t$ region.
\\
On the other hand, in the $C_{f\neq t}=1$ scenario, the signal strengths follows
a (less involved) monotonic behavior.

The parameter dependence just shown will be crucial  in the following analysis to understand  the exclusion potential of the \tqhspp process in the ($C_V,C_f$) plane.

\section{Single top quark and Higgs boson production}

In \,\cite{Biswas:2012bd}, we estimated the sensitivity to an anomalous \tqhspp  production focussing on the events where the Higgs boson decays to photons, and the top quark decays hadronically. The study was based on an integrated luminosity of  60\,fb$^{-1}$  at either 8\,TeV and 14\,TeV. Here the same estimate is repeated for the approximate 50\,fb$^{-1}$ collected by the ATLAS and CMS experiments in the 7 and 8\,TeV LHC runs (in particular $\sim 5$\,fb$^{-1}$ at 7 TeV  plus 
$\sim 20$\,fb$^{-1}$ at 8\,TeV, for each experiment).
Different Higgs-boson and  top-quark decays will be analyzed  in several non-overlapping final states. The sensitivity of all  independent  decay channels of the final state \tqhspp will be thus assessed separately.

\subsection{$H \to \gamma \gamma$}

Here, we first update our previous sensitivity estimate
on the basis of the total luminosity collected so far in the 7 and 8 TeV LHC runs. In addition, we analyze for the first time the sensitivity of the final state where  $t \to b \,\ell \nu$. Throughout the paper, a lepton $\ell$ is assumed to be either an electron or a muon. 

\subsubsection{Final state with a hadronically decaying top quark}

The hadronic top quark decays allow the full reconstruction of the event kinematics. Combinatorics in the jet-to-parton assignment is greatly reduced by the usage of 
$b$-tagging algorithms, and by the requirement that the most forward jet be the one not emerging from the top quark decay. The  irreducible $SM$ backgrounds include top and multi-jet final states, all accompanied by two high-$p_T$ photons, 
\begin{itemize}
\item $pp \to 2\,\gamma+t+j$;
\item $pp \to H(\to \gamma\gamma)+t\bar{t} $;
\item $pp \to 2\,\gamma+\bar{t}\,t$;
\item $pp \to 2\,\gamma+b+3\,j$.
\end{itemize}
We include also the top-pair and Higgs associated production 
$pp \to t\bar{t}H$
as a relevant
background in all channels, taking into account the $Ht \bar{t}$ coupling dependence for consistency (notably giving a null contribution to the background in the  $C_t=0$ case).
Among the reducible backgrounds the dominant contribution comes from $pp \to 2\,\gamma+4\,j$,  where one of the light jets can be mistagged as a $b$-jet.
We assume a $b$-jet identification efficiency of 60\%, with a corresponding fake jet rejection factor of 10 for $c$-jet, 
and 100 for other light jets. 

For the present channel, we retain only the events passing the following selection criteria on the transverse momentum $p_T$ and pseudo rapidity $\eta$ of the final state particles:
\begin{eqnarray}
\begin{tabular}{l}
$p_{T}^{\gamma_1}>40$~GeV, \qquad $p_{T}^{\gamma_2}>30$~GeV, \qquad $p_{T}^{j,b}>25$~GeV, 
\qquad $|\eta_{\gamma,b}|<2.5$,  \qquad $|\eta_{j}|< 4.5$.
\end{tabular}
\label{tag1}
\end{eqnarray}
The above cuts reflect the requirement imposed by the ATLAS and CMS di-photon triggers, and fiducial regions of the detectors for the final state objects.

In all our analysis, we use the following general isolation requirement for  
any pair of objects in the final state
\begin{equation}
\Delta R_{i,j}=\sqrt{ \Delta \eta^2_{i,j}+\Delta \phi^2_{i,j}}> 0.4\;,
\label{tag2}
\end{equation}
with {\it i} and  {\it j} running over all the final particles 
and jets (including $b$-jets),
$ \Delta \eta$ being the rapidity gap, and $ \Delta \phi$  the azimuthal-angle gap between any particle pair. 
\\
For the present channel, 
the highest rapidity light-jet in the final state is required to have $|\eta|>2.5$ and $p_T> 30$  GeV. 
For all other final states  this condition will be somewhat relaxed, thanks to the presence of more characterized final states with  extra charged leptons. 
\\
As in our previous study, we require the top quark to be fully reconstructed in the hadronic mode, with the invariant mass of the remaining 3-jets system (out of which one is a $b$-jet)  peaking at the  top quark mass, within a mass window consistent with the ATLAS and CMS jet energy resolution, i.e. 20 GeV. Then, the invariant mass of the two light jets contributing to the top quark invariant mass, is required to peak at the $W$ mass within a mass window of 15 GeV. Finally,  the invariant mass of the di-photon system is imposed to reconstruct the Higgs mass centered at 125 GeV within a mass resolution of $\pm 3$ GeV.
%
\begin{table}[thp]
\begin{center}
\begin{tabular}{|c|c|c|c|c|c|c|c|c|}
\hline 
$\sqrt{s}=8$ TeV \bs(50 fb$^{-1}$) \es & \multicolumn{3}{c|}{{\it Signal (S)}} & \multicolumn{5}{c|}{{\it Background (B)}} \\
\hline
{\it Cut}  & \bs {\bf $C_t=-1$}\es &  \bs{\bf $C_t=0$}\es & 
\bs{\bf $C_t=1$}\es & \bs$2\gamma\, tj$\es & \bs$2\gamma\, t\bar{t}$\es &  \bs$t\bar{t}H$\es &
\bs$2\gamma\, b\,  3j$\es& { \bs $2\gamma \,4j$ \es} \\
\hline
\bs$2\gamma+b+ (\geq  \;$3\,$j$) \es&  6.4  &  5.1  &  0.18  & 8.2 & 9.2 & 1.6 & 249 & 1263 \\
\bs$|\eta_{j_F}|>2.5$ \& $p_{{T}_{j_F}}>30$~GeV \es& 3.0  & 2.5 & 0.08 & 3.3 & 0.32  & 0.06 & 22  & 116\\
\bs$|M_{bjj}-m_{t}|<20$ GeV\es & 3.0  &  2.4   & 0.08  & 3.3 & 0.20  & 0.02 &  4.5 & 30 \\
\bs$|M_{jj(top)}-M_{W}|<15$ GeV \es& 2.8  &  2.3  & 0.07  & 3.2  & 0.19  & 0.02 & 1.8 & 4.6 \\
\bs$|M_{\gamma\gamma}-m_H|<3$ GeV\es & 2.8  &  2.3  & 0.07 & 0.12 & 0.02 & 0.02 & 0.57  & 0.26 \\
\hline
$S/\sqrt{S+B}$ &  1.4 & 1.3 & 0.07\\
\cline{1-4}
\end{tabular}
\caption{\small \it Number of events passing sequential cuts for the signal \tqaaspp and $SM$ backgrounds  at $\sqrt s=8$ TeV, with integrated luminosity of $50$ fb$^{-1}$, assuming $C_f=C_t$. 
Results correspond to $m_H=125$ GeV. An estimate of the sensitivity of this channel to the signal, given by the significance $S/\sqrt{S+B}$ (with $B$ summed up over all contributions), is shown in the lowest row. 
The contribution of $t\bar{t}H$ to the SM background has been set to zero in the evaluation of $S/\sqrt{S+B}$ at $C_t=0$.} 
\label{tab:eventrate_hggtophad}
\vspace*{-0.3cm}
\end{center}
\end{table}          

The results of the above selection procedure are shown in Table~\ref{tab:eventrate_hggtophad} for $m_H=125$ GeV, $\sqrt s=8$ TeV, and integrated luminosity of 50 fb$^{-1}$. 
The numbers of events passing the sequential cuts  
defined above are reported for the \tqaaspp signal with hadronic top decay, for different $C_t=\pm 1,0$ values (assuming $C_f=C_t$ in \braa),  for the main irreducible  backgrounds, and one reducible background.
The first row refers to the total number of events that pass
the photon- and jet-tagging definition.
 As anticipated, we  include in our analysis of the relevant backgrounds
the $t\bar{t}H$ final state, for all the Higgs decay channels considered.
This will give rise to a mild dependence on  $C_t$ 
of the total background rates  in the range of $C_t$ values considered here. 
The results explicitly shown in the Tables~1-4 for different channels is the total number
of $t\bar{t}H$ background events corresponding to $C_t=\pm 1$. However, in the evaluation of the
significance at $C_t\neq 1$, we consistently tune  the $t\bar{t}H$ event rates  
in the total background. We also estimated the contribution from the reducible background  $2\,\gamma+b\bar{b}+2\,j$, where one of the $b$-quark 
is mistagged as a light jet. The latter is found to be quite small after applying the set of cuts  in Table~\ref{tab:eventrate_hggtophad}, contributing by about 0.01 events  (corresponding to a cross section of $2.0\cdot 10^{-4} $~fb at $8$ TeV).

From these results, we can see the efficiency of the different cuts  applied to enhance the signal-to-background ratio. In particular, the signal rate is mainly affected by the 
forward-jet tag requirement, with a corresponding reduction of roughly a factor 2, and it passes almost unaltered the remaining cuts. The largest contribution to the background comes 
from the $2\,\gamma+b+3\,j \,$ non-resonant final state. This is considerably affected by both the forward-jet cut and the top- and Higgs-resonance requirements. 
The same holds for the reducible $2\,\gamma+4\,j \,$ background\footnote{In this case, after mistagging one of the light jet as a $b$-jet and tagging a forward jet out of the remaining 3 jets, the fake $b$-jet has been combined with the remaining two jets to form the top system.}, which contributes 
by almost 50\% of the $2\,\gamma+b+3\,j \,$ event rate. 
The next background for 
relevance is the single-top production $2\,\gamma+t+j$, while the top-pair channels $2\,\gamma+\bar{t}\,t\,$ and
$H(\to 2\,\gamma)+\bar{t}\,t\,$
contributes to the total rate of background events  negligibly. Finally, in the bottom row of
Table~\ref{tab:eventrate_hggtophad}, we report an  estimate of the sensitivity of this channel to an anomalous $C_t$ value, given by the 
 significance defined as $S/\sqrt{S+B}$, 
for  $C_t=-1,0,1$.

\subsubsection{Final state with a leptonically decaying top quark}

We  now consider the \tqaaspp process, in events where the top quark decays leptonically $t\to b\, \ell\, \nu$, and $\ell = e,\mu$.
The irreducible backgrounds are given by the processes
\begin{itemize}
\item $pp \to 2\,\gamma+t+j$;
\item $pp \to H(\to \gamma\gamma)+t\bar{t} $;
\item $pp \to 2\,\gamma+\bar{t}\,t$;
\item $pp \to 2\,\gamma+W+b+\,j$.
\end{itemize}
The dominant reducible background $pp \to 2\,\gamma \,W \,2\,j$ has also been included where one of the light
jet is mistagged as a $b$-jet.

In addition to the isolation criteria in Eq.(\ref{tag2}), 
the following selection cuts have been applied on the final state particles:
\begin{eqnarray}
\begin{tabular}{l}
$p_{T}^{\gamma}>20$~GeV, \qquad $p_T^{\mu}>20$~GeV, \qquad $p_{T}^{e,j,b}>25$~GeV, 
\qquad $|\eta_{\gamma,l,b}|<2.5$,  \qquad $|\eta_{j}|< 4.5$.
\end{tabular}
\label{tag1b}
\end{eqnarray}
Here, the looser cuts on the photon transverse momentum $p_T^{\gamma}$ 
are allowed by the addition of a  single-lepton trigger.
The most forward jet $j_F$ is required to have a pseudo-rapidity of at  least 1.5.
\begin{table}[thp]
\begin{center}
\begin{tabular}{|c|c|c|c|c|c|c|c|c|}
\hline
$\sqrt{s}=8$ TeV \bs(50 fb$^{-1}$) \es & \multicolumn{3}{c|}{{\it Signal (S)}} & \multicolumn{5}{c|}{{\it Background (B)}} \\
\hline
{\it Cut}  & \bs {\bf $C_t=-1$}\es &  \bs{\bf $C_t=0$}\es & 
\bs{\bf $C_t=1$}\es & \bs$2\gamma\, tj$\es & \bs$2\gamma\, t\bar{t}$\es &  \bs$t\bar{t}H$\es &
\bs$2\gamma\, W\, bj$\es& \bs $2\,\gamma W 2\,j$\es \\
\hline
\bs$2\gamma+\ell+ b+ (\geq  \;$j) \es&  3.01  &  2.35 & 0.08  & 7.0 & 6.5 &  0.8 & 5.0 & 5.57 \\
\bs$|M_{\gamma\gamma}-m_H|<3$ GeV\es & 3.01  &  2.35 & 0.08 & 0.16 & 0.18 & 0.77 & 0.09  & 0.12 \\
\bs$|\eta_{j_{F}}|>1.5$ \es& 2.54  & 2.01 & 0.06 & 0.12 & 0.04  & 0.15 & 0.03  &  0.04 \\
\hline
$S/\sqrt{S+B}$ &  1.5 & 1.3 & 0.09\\
\cline{1-4}
\end{tabular}
\caption{\small 
\it 
Same as in Table \ref{tab:eventrate_hggtophad} 
for the signal \tqaapp, and  $SM$ backgrounds,
but for the  top quark decaying leptonically.} 
\label{tab:eventrate_hggtoplept}
\vspace*{-0.3cm}
\end{center}
\end{table}         
As the dominant backgrounds also feature leptonically decaying top quarks, we do not attempt for a partial top-quark reconstruction. 

In Table~\ref{tab:eventrate_hggtoplept}, we show our results for the event yields for the signal and different  
backgrounds, for the representative $C_t=-1,0,1$ values.
 From the latter results, it can be seen that, by using as simple figure of merit $S/\sqrt{S+B}$, the sensitivity that can be achieved in these class of events is 
comparable to the one where the top quark decays hadronically.

\subsection{$H \to WW^*$ and $H \to \tau \tau$}

 Several combinations of number and charge of leptons are available in the \tqhpp$ \to bWqWW, bWq\,\tau\tau$ final states. We will narrow our study to the ones 
that are of larger potential in terms both of sensitivity to a signal, and in robustness against poorly known backgrounds, i.e. events where at least two $W$'s or two $\tau$'s (or one $W$ and one $\tau$) decay leptonically.
As $BR(H \to \tau\tau)$ is approximately a factor 3 smaller than $BR(H \to WW^*)$, in the SM, and the $\tau$ branching ratio in electrons/muons is larger than $BR(W \to e(\mu)\, \nu)$, 
the final states with multi leptons will in general be sensitive to a mixture of $W$ and $\tau$. 
The $W$'s (both in signal and in  backgrounds) are let to decay 
 only to the leptons of first two generations. On the other hand, the tau decay 
in the $H\to \tau\tau$ signal has been handled using
the in-built interface in the event generator PYTHIA\cite{PYTHIA}\footnote{Though we are working with parton-level jets, for the hadronic tau decay, 
the decay products have been considered as a single jet in our analysis.}. 
We assume  the same  isolation 
criteria for  the leptons coming from  the $W$ and  tau decays.

\subsubsection{Final state with three charged leptons}

The final state in this case is $3\,\ell+b+j$ (with $\ell = e {\rm ~or} ~\mu)$. Two of the three leptons 
have their origin from the decays of two $W$'s or $\tau$'s coming from the Higgs decay, and the additional 
prompt lepton comes from the top semileptonic decay.
The irreducible backgrounds which contribute to this final state are:
\begin{itemize}
\item $pp \to t\bar{t} + W$;
\item $pp \to t\bar{t} + Z$;
\item $pp \to t\bar{t} + H$;
\item $pp \to t + WW + j$;
\item $pp \to t + WZ + j$;
\item $pp \to b + WZ + j$;
\item $pp \to b + WWZ$\,.
\end{itemize}
We impose  the following event selection  criteria
\begin{eqnarray}
\begin{tabular}{l}
$N_{\ell} = 3 \qquad p_{T}^{\ell}>20$~GeV, \qquad $p_{T}^{(j,b)}>20$~GeV \qquad $|\eta_{\ell,b}|<2.5$, \qquad $|\eta_{j}|< 4.5$\, ,
\end{tabular}
\label{tag2b}
\end{eqnarray}
which has a large acceptance to the signal, while keeping large rejection for 
the above backgrounds.
Similarly to the other channels investigated, all final state particles are 
required to be sufficiently isolated by imposing the isolation criteria in 
Eq.(\ref{tag2}).
A rapidity cut $|\eta|>1.5$ on the forward jet has also been applied.

In connection to differences in the backgrounds, it is useful to discriminate among two 
different sets of tri-lepton signatures, depending on the combination of charges and flavors of the leptons. In particular:
\begin{itemize}
 \item {$e^{\pm}e^{\pm}\mu^{\mp}$ and $\mu^{\pm}\mu^{\pm}e^{\mp}$} signatures, 
where out of three leptons, two
leptons have the same charge and same flavor, while the third one has 
opposite charge and different flavor.
This final state is defined here as $\ell_i^{\pm}\ell_i^{\pm}\ell_j^{\mp}+b+jets$ (with $i\neq j$).
\item The complementary signature to the above 3-lepton one, defined as $\ell_i^{\pm}\ell_j^{\pm}\ell_j^{\mp}+b+jets$.
\end{itemize}
\begin{table}[thp]
\begin{center}
\begin{tabular}{|c|c|c|c|c|c|c|c|c|}
\hline
$\sqrt{s}=8$ TeV \bs(50 fb$^{-1}$) \es & \multicolumn{3}{c|}{{\it Signal (S)}} & \multicolumn{5}{c|}{{\it Background (B)}} \\
\hline
{\it Cut}  & \bs {\bf $C_t=-1$}\es &  \bs{\bf $C_t=0$}\es & \bs{\bf $C_t=1$}\es &  {\bf $t\bar{t}W$}  &  {\bf $t\bar{t}Z$} &  {\bf $t\bar{t}H$} & {\bf $tWWj$} & {\it Total} \\
\hline
$\ell_i^{\pm}\ell_i^{\pm}\ell_j^{\mp} b q$ & 0.96  &  0.81  & 0.06  & 3.69  & 0.14  &  1.07  & 0.04 &  4.94 \\
$|\eta^F_{j}|>1.5$ &  0.81  &  0.70  &  0.05 & 0.64 & --  &  0.18  & 0.01 & 0.83 \\
\hline
$S/\sqrt{S+B}$ &  0.63 & 0.57 & 0.05\\
\hline
  & \multicolumn{3}{c|}{{\it Signal (S)}} & \multicolumn{5}{c|}{{\it Backgrounds (B)}}\\
\cline{2-9}
 &  {\bf $C_t=-1$} & {\bf $C_t=0$} & {\bf $C_t=1$}  &  {\bf $t\bar{t}W$}  &  {\bf $t\bar{t}Z$} & {\bf $t\bar{t}H$} & {\bf $tWWj$} & {\it Total} \\
\hline
$\ell_i^{\pm}\ell_j^{\pm}\ell_j^{\mp} b q$ &  3.11  &  2.58    & 0.18  & 12.2  & 43.5  &  3.3 &  0.2  & 59.2 \\ 
$|\eta^F_{j}|>1.5$ & 2.72    &  2.22   & 0.14  & 2.6  &  11.0  & 0.6  & 0.1  & 14.3  \\ 
$M_{\ell^{+}_j \ell^{-}_j}\notin [86.2, 96.2]$ GeV  & 2.16  & 1.76  & 0.11 & 2.0  & 0.2  &  0.4  & -  & 2.6 \\  
\hline
$S/\sqrt{S+B}$ &  0.99 & 0.88 & 0.07 \\
\cline{1-4}
\end{tabular}
\caption{Same as in Table \ref{tab:eventrate_hggtophad}, 
but for the signal \tqhpp, with  $H\to \ell\nu(\nu)\ell\nu(\nu)$ and $t \to b \ell \nu$, and corresponding irreducible backgrounds.}
\label{tab:htoWW_trilep}
\end{center}
\end{table}         

A large fraction of the background is characterized by the emission of a $Z$ vector boson 
in place of the Higgs boson.
Then, in order to suppress the latter backgrounds,  the invariant mass of two opposite-sign (same-flavor) leptons is required to be $M_{\ell^+\ell^-}\notin [86.2,96.2]$.

In Table\;\ref{tab:htoWW_trilep}, we show our results for the event yields, under three different $C_t$ assumptions, and the corresponding background yields. 
The extra background contribution coming from $tWZ, bWZj, bWWZ$ processes, arising in events where the opposite sign leptons share the same flavor, is effectively suppressed to a negligible level once the $Z$ mass veto is applied. The relative contribution of \tqhspp with $H \to W W^*$ and $H \to \tau \tau$ to the tri-lepton signal is approximately 2 to 1.
\subsubsection{Final state with two same-sign leptons}
The final state in this case is $\ell^{\pm }\ell^{\pm} b\,qqq$ (with $\ell = e {\rm ~or} ~\mu)$. In order to achieve the same-sign dilepton production,
one of the charged leptons must arise from the top  decay, while the other from the leptonic decay of the $W($or $\tau)$ that shares the  sign
with the $W$ coming from the top quark. The main backgrounds in this channel are, in order of importance:
\begin{itemize}
\item $pp \to t\bar{t} + W$;
\item $pp \to t\bar{t} + Z$;
\item $pp \to t\bar{t} + H$;
\item $pp \to t + WW + j$;
\item $pp \to t + W + jjj$.
\end{itemize}
The possible acceptance of the adopted event selection to the $bWW3j$ and $bbWW2j$ final states has been investigated, and found to be negligible once normalized
to the 50\,fb$^{-1}$ of integrated luminosity at 8 TeV collisions.

The following event selection has been adopted, according to present experimental analysis in the same-sign dilepton final state\,\cite{Chatrchyan:2013qca}
\begin{eqnarray}
\begin{tabular}{l}
$N_{\ell} = 2 \qquad p_{T}^{\ell}>20$~GeV, \qquad $p_{T}^{(j,b)}>20$~GeV \qquad $|\eta_{\ell,b}|<2.5$, \qquad $|\eta_{j}|< 4.5$\,,
\end{tabular}
\label{tag3}
\end{eqnarray}
with the requirement that the two charged leptons are either both positively or both negatively charged. A rapidity cut $|\eta|>1.5$ on the forward jet has been applied. The corresponding results for the signal and background
events are reported in Table \ref{tab:htoZZ-tjh}.
\begin{table}[htpb]
\begin{center}
\begin{tabular}{|l|c|c|c|c|c|c|c|c|c|}
\hline 
$\sqrt{s}=8$\,TeV \bs(50 fb$^{-1}$) \es & \multicolumn{3}{c|}{{\it Signal (S)}} & \multicolumn{6}{c|}{{\it Backgrounds (B)}}\\
\hline
 $Cut$  &  \!{\bf $C_t = -1$}\! & \!{\bf $C_t = 0$}\! & \!{\bf $C_t = 1$}\!  &  {\bf $t\bar{t}W$} & {\bf $t\bar{t}Z$} & {\bf $t\bar{t}H$} & {\bf $tWWj$} & {\bf $tW\, 3j$} & {\it Total} \\
\hline
$\ell^{\pm }\ell^{\pm} bqqq$ & 7.8   &  6.3  & 0.53   &  45.1  &  3.7   &  8.4  & 0.5  &  0.3      &  57.9  \\
$|\eta^F_{j}|>1.5$~GeV & 6.6    &  5.4  &  0.42  &  11.3   &  0.6   & 1.8   & 0.2    & 0.1    &  13.9   \\
\hline
$S/\sqrt{S+B}$ &  1.5 & 1.3 & 0.11  \\
\cline{1-4}
\end{tabular}
\caption{Same as in Table \ref{tab:eventrate_hggtophad}, for the signal and corresponding 
irreducible background numbers of events, but for final states with two same-sign 
leptons.}
\label{tab:htoZZ-tjh}
\end{center}
\end{table}         

The relative weight of  $H \to W W^*$ and $H \to \tau \tau$ contributions to the \tqhspp final state is again approximately 2 to 1.

\section{Sensitivity to anomalous Yukawa couplings}
After assessing the sensitivity to an anomalous single-top plus Higgs production in different final states, we proceed to study their combined potential, and the implications on the parameter space of the Higgs couplings in the two scenarios highlighted in Section 2. We thus combine the five channels discussed in Section 3, i.e. events with $H \to \gamma \gamma$ and top decaying either leptonically or hadronically, the two trilepton final states (i.e. the set with same-sign same-flavor lepton pairs, and the complementary set), and the same-sign dilepton final states, the latter three channels being   simultaneously sensitive  to $H \to WW^*$ and $H \to \tau \tau$. As previously, we exploit the full LHC luminosity collected at 7 and 8 TeV by ATLAS and CMS. 

Table~\ref{tab:signal-tot} presents a summary of the signal event numbers ($S$), background event numbers ($B$), and corresponding significances $S/\sqrt{S+B}$, for the five channels considered, versus $C_t$ (for the values $C_t=-1.5,-1,0,0.3$) in the universal scaling scenario ($C_f=C_t$), with $C_V\simeq 1$. It is straightforward to see that the signal-rate (and corresponding significance) profile for each channel 
closely follows the signal-strength behavior in Figure~2.
The latter leads to a roughly constant magnitude for both signal rates and significances  for $C_t\lappeq 0$.
The dominant contributions come from the two $\gamma\gamma$ channels 
(with comparable strength), and the two same-sign lepton channel, each contributing by a 1.5-$\sigma$ significance at  
$C_t\sim -1$ (the latter channel having larger event numbers which are compensated
 by  higher background rates).
 Note that in the multi-lepton components, the  $H \to WW^*$ contribution  would 
provide a comparatively good  signal rate  {\it closer} to the $C_t\sim 0$ region with respect to the $\gamma\gamma$ channels (cf. Figure~2).  The latter behavior is actually compensated by the $H \to \tau \tau$ component of the multi-lepton channels  (giving on average 30\% of the total multi-lepton rates), which rapidly drops for $C_t\to  0$.
Altogether, the total significance    
approaches the 3-$\sigma$ level in all the  $C_t\lappeq 0$ region, for $C_V\simeq 1$ 
(as shown in the last column of 
Table~\ref{tab:signal-tot}).

\begin{table}[htpb]
\begin{center}
\begin{tabular}{|l|ccccc|c|}
\hline 
 & \multicolumn{5}{c|}{{\it Channels}} & \multicolumn{1}{c|}{{\it Total}}  \\
\hline
 Process  & $\gamma \gamma bqq q'$ & $\gamma\gamma b\ell q'$ & $\ell_i^{\pm}\ell_i^{\pm}\ell_j^{\mp}b q'$ & $ \ell_i^{\pm}\ell_j^{\pm}\ell_j^{\mp} b q'$ & $\ell^{\pm }\ell^{\pm} bqq q'$  & $ $\\
\hline
\hline
${\bf C_t = -1.5}$ & \multicolumn{5}{c|}{} & \multicolumn{1}{c|}{}\\
\hline
$S$ & 2.6 & 2.4 & 0.91 & 2.4 & 3.6 & 11.9  \\
$B$ &          
1.04 & 0.73 & 0.86 &  2.7 &   14.3 & 19.6 \\
$S/\sqrt{S+B}$ & 
1.4    & 1.4    & 0.68    & 1.1    & 0.85    &
2.5
\\
\hline
\hline
${\bf C_t = -1}$ & \multicolumn{5}{c|}{} & \multicolumn{1}{c|}{}\\
\hline
$S$ & 2.8 & 2.5 & 0.81 & 2.2 & 6.6 & 14.9  \\
$B$ & 1.02 & 0.60 &  0.83 &  2.6 &   14.0 & 19.1 \\
$S/\sqrt{S+B}$ & 
1.4    & 1.4    & 0.63    & 1.0    & 1.5    &
2.8
\\
\hline
\hline
${\bf C_t = 0}$ & \multicolumn{5}{c|}{} & \multicolumn{1}{c|}{}\\
\hline
$S$ & 2.3 & 2.0 & 0.70 & 1.8 & 5.4 & 12.2 \\
$B$             & 
0.97 & 0.23 & 0.65 &  2.2  &  12.2 & 16.2 \\
$S/\sqrt{S+B}$ &
1.3    & 1.3    & 0.60    & 0.90    & 1.3    &
2.5
\\
\hline
\hline
${\bf C_t = 0.3}$ & \multicolumn{5}{c|}{} & \multicolumn{1}{c|}{}\\
\hline
$S$ & 1.0 & 0.80 & 0.33 & 0.84 & 2.5 & 5.5  \\ 
$B$ & 0.98  & 0.29 &  0.70 &  2.3  &  12.7 & 17.0 \\
$S/\sqrt{S+B}$ & 
0.71    & 0.77    & 0.33    & 0.47    & 0.64    &
1.4
\\
\hline
\end{tabular}
\caption{Event rates ($S$) for the signal \tqhpp, for different $C_t$ values and $C_f=C_t$ (at $C_V\simeq 1$), in the five different final states corresponding to the decays $H \to 
\gamma \gamma, ~WW^*,~ \tau \tau$, with integrated luminosity of $50$ fb$^{-1}$ at 8 TeV. 
The corresponding background rates ($B$) and related significances are also detailed. The total significance (last column) is obtained by summing up in quadrature individual significances.
} 
\label{tab:signal-tot}
\end{center}
\end{table} 

\begin{figure*}[t]
\begin{center}
\includegraphics[width=1\textwidth]{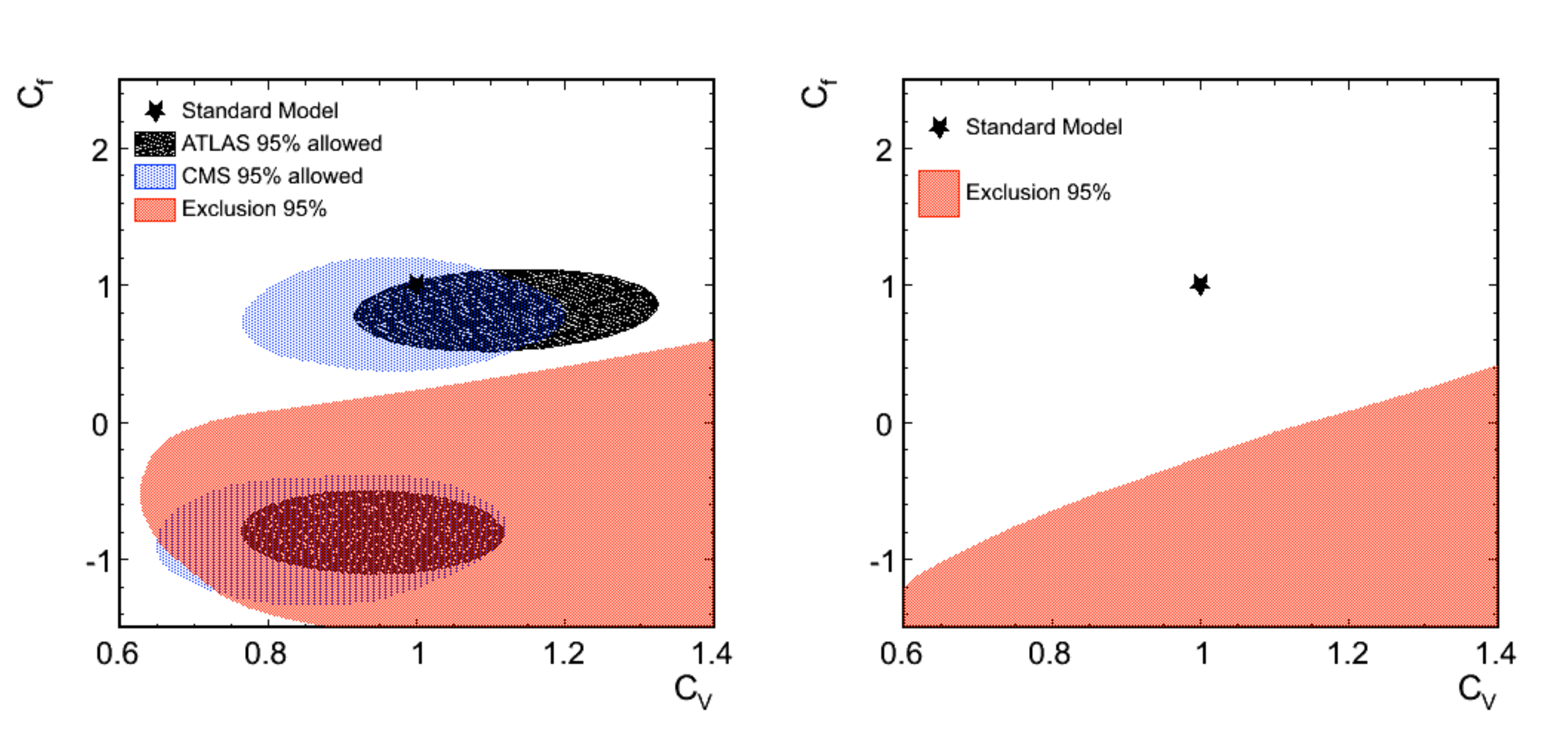}
\vspace{-0.3cm}
\caption{The left plot shows the $(C_V,C_f)$ plane with the currently allowed parameter regions by the ATLAS (dark shaded oval) and CMS (light shaded oval) experiments \cite{atlas-11,cms-11}, in the universal $C_f$ scaling assumption. The red dashed area corresponds to the 95\% C.L. exclusion contours that can be achieved by searching for anomalous  \tqhspp production with $H \to \gamma\gamma,WW^*,\tau \tau$, in 50\,fb$^{-1}$ of LHC data
at 8 TeV. The right plot shows the 95\% exclusion contour
for a free  $C_t$  with $C_{f\neq t} = 1$. 
 Negative-sign regions for the $C_f$ scaling factor can be excluded in both scenarios.}
\label{fig:excl}
\end{center}
\end{figure*}

The five different channels are statistically independent by construction. No systematic uncertainties are estimated for the time being. 
Then, we obtained  the 95\% confidence-level (C.L.) upper limits on an anomalous  cross section for \tqhspp by building a likelihood that is the product of the likelihoods for all channels. We  report the results of our analysis in the $(C_V,C_f)$ plane, in the two hypotheses where either a universal rescaling of the Higgs couplings to fermions  is assumed ($C_f = C_t$), or  $C_t$ is a free parameter, and the remaining scaling factors are fixed to their SM value ($C_{f\neq t} = 1$). Figure~\ref{fig:excl} shows the exclusion contour in these two assumptions, along with the regions currently favored by  the fits to present Higgs data in the universal scaling scenario \cite{atlas-11,cms-11}. 
In both scenarios, the potential of the combination of the five channels considered here for the exclusion of the  $C_t  \lappeq 0$ region is quite remarkable.
In particular, in the universal $C_f$ scaling scenario, one can completely cover at this stage  all the 
 negative $C_f$ space compatible with global fits. 
\\
It is also interesting to note a partial complementarity with the potential of the 
$H\to \bar b b$ 
channel, performed in  \cite{Farina:2012xp}. In the universal rescaling hypothesis, the main sensitivity in the region 
$|C_t| \lappeq 1$ comes from the decays of the Higgs into bosons, as  of the branching ratios of the Higgs to fermions (notably $BR_{bb}$) is depleted. 
\\
Conversely, assuming  $C_{f\neq t} = 1$, there is no large enhancement in the branching ratio of the Higgs to $WW^*,\tau \tau$ (and only a moderate increase in the 
$H\to \gamma\gamma$ rate, cf. Figure~3), thus resulting in lower sensitivities. In both scenarios, the negative minimum lies well within the contours of the excluded region.

The parton level analysis  performed here thus shows that studying the \tqhspp production in bosonic Higgs decays is the most sensitive approach for constraining an anomalous top-Higgs Yukawa coupling in the negative $C_f$ region.

\section{Conclusions}
The sensitivity to an anomalous top Yukawa couplings 
of the single top and Higgs associated production $pp\to t q H$ has been assessed  with the Higgs and the top quark decaying to the most robust final states 
(that is either resonant photon pairs or multi-lepton final states)  
for a luminosity of $50~{\rm fb}^{-1}$ at 8 TeV collisions,  roughly corresponding to the present data set collected by  the ATLAS and CMS experiments.
In particular, we extended our previous study in \cite{Biswas:2012bd} by both 
complementing the  $H\to \gamma \gamma$ analysis made for a hadronic
top decay  with a semi-leptonic top decay, and  including multi-lepton final states originating by a Higgs decaying into  $WW^*$ and $\tau\tau$ pairs.
Correspondingly, apart from the two-photon channels, we analyzed the three-lepton and two-same-sign-lepton signatures arising from  the leptonic and semi-leptonic Higgs decay modes through $WW^*$ and $\tau\tau$ pairs combined with the semi-leptonic top decays.
We presented  a parton-level analysis, and  assumed the dominance of the corresponding {\it irreducible} backgrounds, which, for multi-photon and multi-lepton final states, is in general a reliable hypothesis. We anyhow also included a few  backgrounds among the dominant reducible ones for the 2-photon final states.

Remarkably, we find that combining the aforementioned 
multiple final states,
on the basis of the $50~{\rm fb}^{-1}$ statistics already collected at 7 and 8 TeV collisions, an excess in the  $pp\to t q H$ cross section 
could  signal the presence of an anomalous top-Yukawa coupling with flipped
sign with respect to the SM value.
 In the universal scaling hypothesis for the Yukawa sector, 
the non observation of such an excess could anyhow exclude at 95\% of C.L.   the reversed-sign region, in the ($C_V,C_f$) plane presently allowed by the combination of all Higgs measurements in other channels \cite{atlas-11,cms-11}. In a scenario where the anomalous behavior is restricted to the top quark, the $pp\to t q H$ can still exclude values $C_t\lappeq -0.2$ for $C_V\simeq1$.
The potential of the considered signatures shows some complementarity with the $pp\to t q H$ potential in the $H\to \bar bb$ channel~\cite{Farina:2012xp}.
\\
We then urge the ATLAS and CMS collaborations to perform our analysis in a more realistic environment,
in order to either confirm or exclude 
the negative-sign region of the top Yukawa scaling factor, which is presently 
allowed by the  ATLAS and CMS Higgs-coupling fits. 


\vspace{1cm}
\vbox{
\noindent{\bf Acknowledgments}  \\
\vspace{-0.2cm}
\noindent \\
S.B. would like to thank the cluster computing facilities of RECAPP, Harish-Chandra Research Institute.
F.M. acknowledges the support of the ``Rita Levi Montalcini" program at Ministero Istruzione Universita` e Ricerca.
E.G.  thanks the PH-TH division of CERN for its kind hospitality
during the preparation of this work. This work was supported by the 
Estonian Science Foundation grant MTT60, by the recurrent financing 
SF0690030s09 project and by the European Union through the European Regional Development Fund.
}


\end{document}